\documentclass[]{aa}
\usepackage{epsfig}

\usepackage{graphicx} 
\usepackage{array}

\usepackage{tabularx}
\usepackage{longtable}

\usepackage{natbib,twoopt}

\begin{document}
\newcommand{\kms}{km~s$^{-1}$}
\newcommand{\Msun}{M$_{\odot}$}
\newcommand{\Teff}{$T_{\rm eff}$}
\newcommand{\FeH}{[Fe/H]}
\newcommand{\Cir}{$^{12}$C / $^{13}$C}
\newcommand{\bacchus}{{\footnotesize BACCHUS}}
\newcommand{\vald}{{\footnotesize VALD}}

\title{Binary properties of CH and Carbon-Enhanced Metal-Poor stars}
\author{
A. Jorissen\inst{1}
\and 
S. Van Eck\inst{1}
\and
H. Van Winckel\inst{2}
\and
T. Merle\inst{1}
\and
H.M.J. Boffin\inst{3}
\and
J. Andersen\inst{4}
\and
B. Nordstr\"{o}m\inst{4}
\and 
S. Udry\inst{5}
\and
T. Masseron\inst{6}
\and
L. Lenaerts\inst{1}
\and
C. Waelkens\inst{2}
}
\institute{Institut d'Astronomie et d'Astrophysique, Universit\'e Libre de Bruxelles, Campus Plaine C.P. 226, Boulevard du Triomphe, B-1050 Bruxelles, Belgium
\and
Instituut voor Sterrenkunde, Katholieke Universiteit Leuven, Celestijnenlaan  200D, 3001 Leuven, Belgium
\and
ESO, Alonso de C\'ordova 3107, Casilla 19001, Santiago, Chile
%              \email{hboffin@eso.org}
\and
Niels Bohr Institute, University of Copenhagen, Denmark
\and 
Observatoire de Gen\`eve, Universit\'e de Gen\`eve, Suisse
\and
Institute of Astronomy, Cambridge University
}

\date{Received; Accepted}

\abstract{The HERMES spectrograph installed on the 1.2-m~Mercator
telescope has been used to monitor the radial velocity of 13 low-metallicity carbon stars, among 
which 7 Carbon-Enhanced Metal-Poor (CEMP) stars and 6 CH stars
(including HIP~53522, a new member to the family, as revealed by a detailed abundance study). 
All stars but one  show clear evidence for
binarity. New orbits are obtained for 8 systems. The sample covers an extended range in orbital periods, extending from 3.4~d
(for the dwarf carbon star HE 0024-2523) to
about 54~yr (for the CH star HD~26, the longest known among barium, CH
and extrinsic S stars). 
Three systems exhibit low-amplitude velocity variations with periods close to 1-yr  
superimposed on a long-term trend.
In the absence of an accurate photometric monitoring of these systems, it is not clear yet
whether these variations are the signature of a very low-mass companion, or of regular envelope pulsations.
The period -- eccentricity ($P - e$) 
diagram for the 40   low-metallicity carbon stars with orbits now available shows no difference between CH and CEMP-s stars (the latter corresponding to those
CEMP stars enriched in s-process elements, as are CH stars). We suggest that they must be considered as one and the same family and that their different names only stem from historical reasons. Indeed, these two families have as well very similar mass-function distributions, corresponding 
to companions with masses in the range
0.5~--~0.7~M$_{\odot}$, indicative of white-dwarf companions, adopting 0.8~--~0.9~M$_{\odot}$ for the primary
component. This result confirms that CH and CEMP-s stars obey the same
mass-transfer scenario as their higher-metallicity analogs, the barium stars.
The $P - e$ diagrams of barium, CH and CEMP-s stars are indeed very similar. They  reveal two different
groups of systems: one with  short orbital periods ($P < 1000$~d) and
mostly circular or almost circular orbits, and another with longer-period and
eccentric ($e > 0.1$) orbits. These two groups either trace different evolutionary channels during the mass-transfer 
episode responsible for the chemical peculiarities of the Ba/CH/CEMP-s stars, or result from the operation 
of tidal circularisation in a more recent past, when the current giant star was ascending the first giant branch.
} 
% 5 {} token are mandatory


\maketitle

\section{Introduction}
\label{Introduction}

One of the most surprising results of the large wide-field spectroscopic surveys for metal-poor stars in our Galaxy \citep[e.g., the HK, HES, and SDSS/SEGUE surveys;][ respectively]{1992AJ....103.1987B,2001A&A...375..366C,2009AJ....137.4377Y} is the high frequency of carbon-enhanced stars ([C/Fe]~$> 1.0$, hereafter CEMP stars) among metal-poor stars. 
The results of the above surveys indicate that they represent 20\% of stars with [Fe/H]~$< -2$ \citep{2003RvMA...16..191C,2005ApJ...625..825L,2006ApJ...652L..37L,2013AJ....145...13A,2013ApJ...762...26Y,2013AJ....146..132L}, but that 
frequency is increasing rapidly as metallicity decreases \citep{2014ApJ...797...21P}. This finding has prompted a number of high-resolution, high signal-to-noise studies aiming at understanding the origin of the abundance anomalies in these objects. The carbon-enhancement phenomenon appears in stars that exhibit four different heavy-element abundance patterns \citep{2005ARA&A..43..531B,2010A&A...509A..93M}:
\begin{itemize}
\item CEMP-s: [C/Fe]$>+1.0$, [Ba/Fe]$>+1.0$, and [Ba/Eu]$>+0.5$. This most numerous class is characterised by enrichments of neutron-capture elements,
with an abundance pattern compatible with the operation of the s-process in asymptotic giant branch (AGB) stars. 
After initial studies of their binary frequency by \citet{2001AJ....122.1545P}, \citet{2003ApJ...592..504S}, and \citet{2005ApJ...625..825L},
 \citet{2014MNRAS.441.1217S} finally demonstrated on a large sample that these stars are all members of binary systems. Thus, it is now established that these CEMP-s stars -- along with the closely-related classical CH stars \citep{1942ApJ....96..101K} -- are members of wide binary systems, where the former primary star transferred material during its AGB phase onto the presently observable companion \citep{1984ApJ...280L..31M,1990ApJ...352..709M}. 
\item CEMP-rs: [C/Fe]$>+1.0$ and $+0.5>$~[Ba/Eu]~$>0.0$. This other class of CEMP stars, exhibiting large overabundances of elements produced by the s-process and of elements traditionally related to the r-process, was discovered by \citet{1997A&A...317L..63B} and \citet{2000A&A...353..557H}. A number of these stars exhibit radial-velocity variations \citep{2004A&A...413.1073S,2005A&A...429.1031B}. Here again the companion could be
responsible for the peculiar abundance pattern \citep[e.g.,][]{2010A&A...509A..93M,2011MNRAS.418..284B}.
\item CEMP-no: [C/Fe]$>+1.0$ and [Ba/Fe]$\sim 0$. \citet{2014MNRAS.441.1217S} found these CEMP stars with no enhancements in their neutron-capture-element abundances  \citep[hereafter CEMP-no;][]{2002ApJ...567.1166A} to have normal binary frequency.
Consequently, a mass-transfer scenario comparable to that occurring in CEMP-s (and possibly in CEMP-rs) stars has not operated here.
\item CEMP-r: [C/Fe]$>+1.0$ and [Eu/Fe]$>+1.0$. This is the signature of a highly r-process-enhanced CEMP star, so far  observed in only one object \citep[CS22892-052;][]{2003ApJ...591..936S}.
\end{itemize}

This paper focuses on the classes of CH and CEMP-(r)s stars, which may be considered as the low-metallicity analogs of barium and extrinsic S stars \citep{1951ApJ...114..473B,1988ApJ...333..219S,1988A&A...198..187J}. \citet{1942ApJ....96..101K} coined the term 'CH stars'
to refer to high-velocity carbon stars that exhibit very strong G
bands due to the CH molecule and otherwise weak metallic lines. The few spectra available at that time suggested high luminosities,
i.e., giant stars. The nomenclature was extended to
objects of lower luminosity by \citet{1974ApJ...194...95B}, who found subgiants with
similar chemical peculiarities (thus called 'subgiant CH stars'). The criteria defining CH and CEMP-s stars, as reviewed above \citep[see as well][]{2005ARA&A..43..531B}, do not allow us to distinguish these stars on spectroscopic or abundance grounds. Their different name appears to be the result of different discovery channels. It is one of the aims of the present study to check that this conclusion of indistinguishability holds true for their orbital properties as well. Indeed, a key property of these two families is their binary nature, as we now discuss. 

An important question concerning the nature of CEMP stars and their various counterparts is 
the astrophysical origin of the carbon excess which is observed in the atmosphere of these
objects. One established scenario is carbon production by
nucleosynthesis in an asymptotic giant branch (AGB) star. 
It is usually assumed that the AGB star
responsible for carbon production was once the primary of a binary
system. The carbon and s-process elements produced in the interior of the AGB star and dredged to its atmosphere were subsequently transferred to a companion. Observational support for this scenario is provided by the binary nature of these objects, as established by \citet{1984ApJ...280L..31M} and \citet{1990ApJ...352..709M} for a major fraction of CH stars, and by \citet{2001AJ....122.1545P} and \citet{2005ApJ...625..825L} for a few CEMP-s stars. 

The mass-transfer scenario responsible for the pollution of the atmosphere of those
stars with s-process material has been the topic of many studies \citep[e.g.,][]{1988A&A...205..155B,2008A&A...480..797B,2010A&A...523A..10I}, which rely on the properties of the  period -- eccentricity   ($P - e$) diagram as well as on the mass-function distribution \citep{1990ApJ...352..709M,1998A&A...332..877J}. For this reason, it is important to collect as many orbital data as possible for CH and CEMP-s stars, so as to build their $P - e$ 
diagram and see whether it bears similarities with that of barium and extrinsic S stars \citep{1998A&A...332..877J,2013EAS....64..163G}, as well as to compare CH and CEMP-s binaries. Since some CEMP-s stars have quite low a metallicity, it is of interest to see whether this low metallicity impacts the $P - e$ diagram (through for instance the change in the stellar size resulting 
from the shift of the low-metallicity giant branches towards bluer colours). 
For CEMP-s stars, very few orbital elements are currently known though \citep[see Table~4 of][]{2005ApJ...625..825L}. Thanks to the new orbits presented 
in this and the paper by \cite{2015A&A...NNN..MMMH}, the sample of orbital elements for CEMP binaries has doubled. The $P - e$ diagram drawn from this extended sample  should represent a benchmark for theoretical mass-transfer studies like those of \citet{2013A&A...552A..26A,2015A&A...576A.118A,2015A&A...581A..22A}.

The paper is organized as follows. Sect.~\ref{Sect:HERMES} describes the  technical details of our radial-velocity monitoring. The stellar sample is described in Sect.~\ref{Sect:sample}. The orbital elements  are presented in Sect.~\ref{Sect:results}. The mass distribution and the $P - e$ diagram are discussed in Sect.~\ref{Sect:elogP}. Sect.~\ref{Sect:sprocess} discusses how to correctly investigate a possible correlation between orbital period and s-process overabundance for these mass-transfer binaries. Conclusions are drawn in Sect.~\ref{Sect:conclusions}.

\section{Radial-velocity monitoring with the HERMES spectrograph}
\label{Sect:HERMES} 

The radial-velocity (RV) monitoring was performed with the  HERMES spectrograph attached to the 1.2-m Mercator telescope 
from the Katholieke Universiteit Leuven, installed at the Roque 
de los Muchachos Observatory (La Palma, Spain). The spectrograph began regular science operations in June 2009, and is fully described 
in \citet{2011A&A...526A..69R}.  The fibre-fed HERMES spectrograph is designed to 
be optimised both in stability as well as in efficiency. It samples the whole optical range from 
380 to 900~nm in one shot, with a spectral resolution of about $86\,000$ for the high-resolution science fibre. This fibre has a 2.5 arcsec aperture on the 
sky and the high resolution is reached by mimicking a narrow slit using a two-sliced image 
slicer. 

The HERMES/Mercator combination is precious because it guarantees
regular telescope time. This is needed for our monitoring programme and
the operational agreement reached by all consortium partners (KULeuven,
Universit\'e libre de Bruxelles, Royal Observatory of Belgium,
Landessternwarnte Tautenburg) is optimised to allow efficient long-term
monitoring, which is mandatory for this programme. On average, 250
nights per year are available for the monitoring (spread equally on HERMES-consortium time, and on KULeuven time), 
and the observation
sampling is adapted to the known variation time scale. During these
nights, about 300 target stars are monitored, addressing several science
cases \citep{2010MmSAI..81.1022V,2013EAS....64..163G}.

\begin{figure}
\includegraphics[width=9cm]{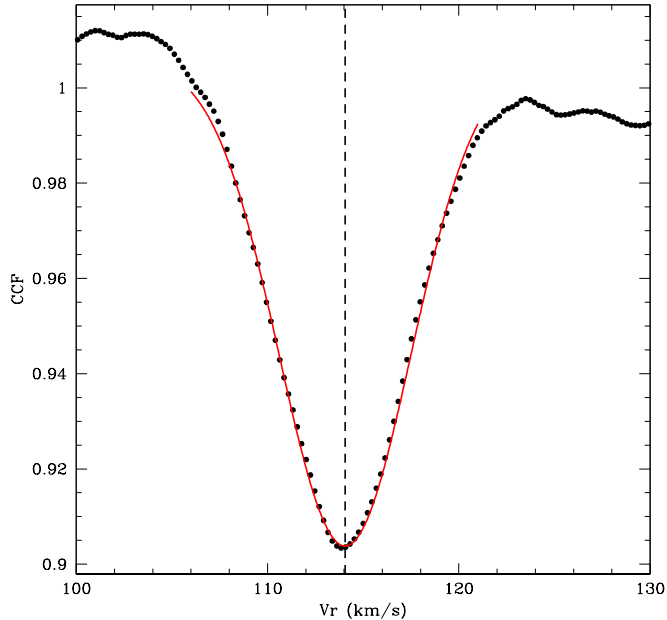}
\caption{\label{Fig:CCF}
A typical CCF for the CEMP-s star HE 1120-2122 on HJD 2456804.428, based on a spectrum with an exposure time of 1800~sec. The Gaussian fit  of the CCF is the thin solid line, restricted to the CCF core to reduce as much as possible the impact of line asymmetries. The vertical line 
marks the star radial velocity as derived from the Gaussian fit. 
}
\end{figure}

A Python-based pipeline extracts a wavelength-calibrated and a
cosmic-ray cleaned spectrum. A separate routine is used for measuring
RVs, by means of a cross-correlation with a spectral mask constructed
from a carbon-star spectrum.  A restricted region, covering the range
478.11 -- 653.56~nm (orders 55 -- 74) and containing 2103 useful
spectral lines, was used to derive the RV, in order to avoid telluric
lines on the red end, and crowded and poorly-exposed spectra on the blue end. A
spectrum with a signal-to-noise ratio of 15 in $V$ is usually sufficient to
obtain a cross-correlation function (CCF) with a well-pronounced
extremum.  An example of CCF  is shown in Fig.~\ref{Fig:CCF} for a
spectrum with an average signal-to-noise ratio of 17 in the relevant
spectral region. 

\begin{figure}
\includegraphics[width=9cm]{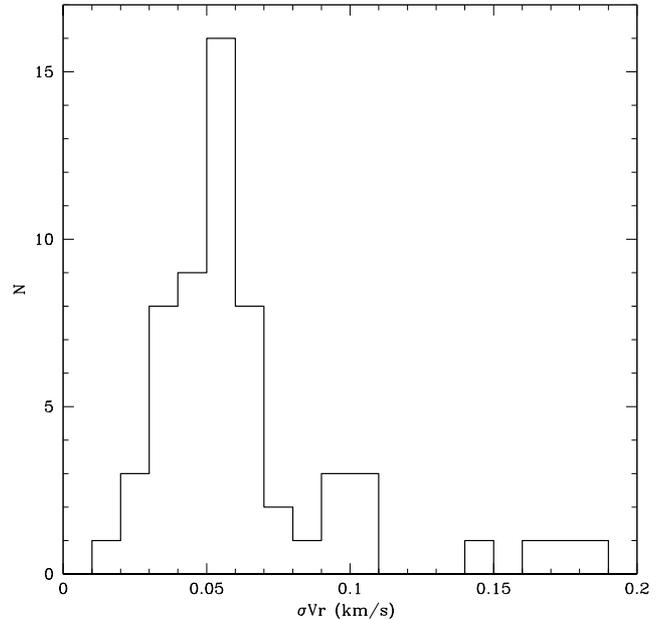}
\caption{\label{Fig:RVstandard}
Histogram of the standard deviations of the velocity data for radial-velocity standard stars with
more than 5 measurements spread over 4.5 years of monitoring.
}
\end{figure}

A Gaussian fit is performed on the core of the CCF (to reduce as much as
possible the impact of line asymmetries), with an internal precision of
a few m/s on the position of the CCF centre, small enough with respect
to external uncertainty factors. The most important of these is caused
by the pressure fluctuations occurring during the night in the
spectrograph room \citep[see Fig.~9 of][]{2011A&A...526A..69R}. The
exact impact of the pressure drift on the velocity measurement depends
on the time elapsed between the science exposure and the arc spectrum
used for wavelength calibration. The long-term accuracy (over years) of
the radial velocities may be estimated from the stability of the RV
standard stars monitored along with the science targets. These standard
stars are taken from the list of \citet{1999ASPC..185..367U}, available
at {\tt http://obswww.unige.ch/$\sim$udry/std/std.html}.
Fig.~\ref{Fig:RVstandard} shows the distribution of standard deviations for 58
radial-velocity standard stars with more than
5 measurements after 4.5 years of monitoring.  The distribution peaks at
$\sigma(Vr)=55$~m/s, which may thus be adopted as the typical uncertainty on the
radial velocities over the long term. The RV standard stars have been
used as well to tie the HERMES RVs to the IAU standard system.

In some cases, older measurements from CORAVEL \citep{1979VA.....23..279B} or from \citet{1990ApJ...352..709M} have been added to the HERMES measurements in order to extend the time span. 
In the latter case (Dominion Astrophysical Observatory velocity measurements), 
a zero-point  correction of $-0.46$~km~s$^{-1}$ has been applied to bring them into the IAU/HERMES system \citep{1998A&A...332..877J}.
The CORAVEL measurements have been tied to the IAU system following the prescription given by \citet{1999ASPC..185..367U}. More specifically for HD~26 (belonging to the sample that will be described in the next section), which is an extreme case given its low velocity of -215~km~s$^{-1}$, the velocities from the northern CORAVEL have first been brought into the IAU system by adding 0.7~\kms\ \citep[taken from the upper left panel of Fig.~2 of][ since the offset due to the extreme velocity overwhelms that due to the moderate $B-V$ index of 1.16]{1999ASPC..185..367U}. Then, the southern CORAVEL measurements have been corrected by -1.3~\kms; the southern measurements are so numerous that the necessary offset can be easily estimated from the shift needed to superimpose the southern and northern sequences.

\section{Stellar sample}
\label{Sect:sample}

The stellar sample, containing 7 CEMP-(r)s and 6 CH stars, 
is listed in Table~\ref{Tab:sample}. It was selected so as to avoid stars 
already monitored by \citet{2005ApJ...625..825L}, and to ensure that the targets are not too far south and not too faint to be observable with HERMES/Mercator.
The spectral classification is from  \citet{2005MNRAS.359..531G}, \citet{2010MNRAS.402.1111G} and \citet{2010A&A...509A..93M}, who provide detailed abundances for the CEMP stars, indicating that they indeed belong to the CEMP-s subclass. 
The star HD~145777 has been included as well. It was initially 
classified as an R3 star by \citet{1940BHarO.913....7M} and R4 by \citet{1944ApJ....99..145S}, but later reclassified as CH by \citet{1956VA......2.1428B}.
The same holds true for HD~187216, HIP~53522 and HIP~53832 \citep{1958AJ.....63..477V,1985AJ.....90.2244H}. The latter two stars
are the topic of a recent paper \citep{2015ApJ..NNN..MMMS} which classifies them as CH-like \citep[a class related to the CH stars, but which lacks the large spatial velocities of the CH stars;][]{1975PASJ...27..325Y}.

Moreover, an abundance analysis of HIP~53522 is presented in Appendix~\ref{Sect:appendix}. Its metallicity was found to be no lower than [Fe/H]$ = -0.5$, a rather moderate value among CH stars, but the s-process elements are clearly enhanced by about 1.4~dex, a result confirmed by the recent analysis by \citet{2015ApJ..NNN..MMMS}.
The star HIP~53832 is very similar to HIP~53522: an abundance analysis by \citet{2009A&A...508..909Z} has confirmed its CH nature, with [Fe/H]$ = -0.77$ and an average s-process abundance of 1.26~dex.

\begin{table*}
\caption{\label{Tab:sample}
Properties of the stellar sample.
$N_{\rm obs}$ is the number of radial-velocity observations \citep[including older measurements from][]{1990ApJ...352..709M}, $\Delta t$ is their time span, $\langle$V$\rangle$ is the average velocity and $\sigma$ its standard deviation, $\sigma/\epsilon$ is the ratio between the standard deviation and the measurement error (estimated to be 55~m~s$^{-1}$). In column SB/ORB, the systems with close to 1-yr variations superimposed on the long-term orbit are flagged with the label  "jitter?", the question mark indicating that a phenomenon intrinsic to the giant's enveloppe ("jitter") is not the only possible cause of these variations, a second, low-mass companion could be invoked as well. 
}
\begin{tabular}{l c c c c c r c r l}% neuf colonnes
\hline\hline
\multicolumn{1}{c}{Name} & Sp. type & Ref. & [Fe/H] & $N_{\rm obs}$ & $\Delta t$  &  \multicolumn{1}{c}{$\langle$V$\rangle$} & \multicolumn{1}{c}{$\sigma$}  & $\sigma/\epsilon$ & SB/ORB \\ 
    & & & & & (d) & (km/s) & (km/s)\\
\hline
HD 26 & CH & 11 & -1.3 &  110 & 11999 & -213.14 & 1.36 & 25 & ORB:\\
HD 76396 & CH  & 6 & - & 68 & 13217 & -61.73 & 1.67 & 30 & ORB+jitter?\\
HD 145777 & CH & 3 & - & 32 & 2215 & 18.43 & 0.83 & 15 & SB\\
HD 187216 & CH & 9 & -2.5 & 41 & 12373 & -115.08 & 4.49 & 82 & SB\\
HIP 53522 & CH(-like) & 6,10,12 & -0.5 & 57 & 1816 & 32.49 & 5.54 & 100 & ORB\\
HIP 53832 & CH(-like) & 6,7,12 & -0.8 & 38 & 1151 & -2.28 & 7.36 & 134 & ORB\\
HE 0017+0055 & CEMP-rs & 1,5 & -2.7 & 25 & 1581 & -79.84 & 0.52 & 9.5 & SB+jitter?\\
HE 0111-1346 & CEMP-s & 1,5 & -1.9 & 52 & 1542 & 44.71 & 6.45 & 117 & ORB\\
HE 0457-1805 &  CEMP-s & 1,5 & -1.5 & 29 & 1894 & 66.76 & 4.24 & 77 & ORB\\
HE 0507-1653 &  CEMP-s & 2,5 & -1.8 & 30 & 1495 & 348.41 & 4.90 & 89 & ORB\\
HE 1120-2122 & CEMP-s & 4 & - & 40 & 1948 & 108.6 & 3.79 & 69 & ORB+jitter?\\
HE 1429-0551 & CEMP-s  & 1,2 & -2.5 & 39 & 1930 & -44.51 & 0.19 & 3.4 & SB?\\
HE 2144-1832 & CEMP-s & 1 & - & 22 & 2170 & 141.79 & 0.60 & 11 & SB\\
\hline
\end{tabular}

References: (1) \citet{2005MNRAS.359..531G} (2) \citet{2010A&A...509A..93M} (3) \citet{1956VA......2.1428B} (4) \citet{2010MNRAS.402.1111G}
(5) \citet{2011AJ....141..102K} (6) \citet{1985AJ.....90.2244H} (7) \citet{2009A&A...508..909Z} 
(8) \citet{1996ApJS..105..419B} (9) \citet{1994A&A...290..148K} (10) This work (11) \citet{2003A&A...404..291V} (12)  \citet{2015ApJ..NNN..MMMS}
\end{table*}

\section{Results}
\label{Sect:results}

\subsection{Synopsis}

A synopsis of the results of the radial-velocity monitoring is presented in Table~\ref{Tab:sample}. The stars may be divided in four different groups:
\begin{itemize}
\item {\bf Star with no clear evidence for binary motion (SB?).} In our sample, only HE~1429-0551  belongs to that category (Fig.~\ref{Fig:Vr1}; for the sake of clarity, all figures with radial-velocity curves are collected in Appendix~\ref{Sect:RV-fig}). Although HE~1429-0551 exhibits radial-velocity variations above the 3$\sigma$ level, and would thus formally qualify for being flagged as spectroscopic binary, inspection of its radial-velocity data  does not reveal any clear trend that could be ascribed to an orbital motion. More observations are thus needed to issue a definite statement about the binarity of that star. 
\item {\bf Stars with clear evidence for binary motion but no orbit available yet (SB).}  Three stars belong to that category 
(HD 145777, HD 187216,  HE 2144-1832; Figs.~\ref{Fig:Vr2}--\ref{Fig:Vr4})  plus a fourth one (HE~0017+0055; Fig.~\ref{Fig:jitter1}) being special in that it shows about 1-yr variations on top of the long-term trend.
\item {\bf Stars with orbits available  (ORB).} Orbits are already available for 4 among the 7 CEMP stars, and for 4 among 6 CH stars (Figs.~\ref{Fig:orbit1}--\ref{Fig:orbit6}). Two among these orbits show about 1-yr-period variations superimposed on the long-term orbit (Figs.~\ref{Fig:jitter2}--\ref{Fig:jitter3}). Orbital elements are listed in Table~\ref{Tab:orbits}. 
\item {\bf Stars with short-period, low-amplitude variations superimposed on the long-term (SB+jitter) or orbital variations (ORB+jitter).} This unexpected behaviour is encountered for the three systems HE~0017+0055, HE~1120-2122, and HD~76396 (Figs.~\ref{Fig:jitter1}--\ref{Fig:jitter3}),  and will be discussed in Sect.~\ref{Sect:short-term}. 
\end{itemize}

In summary, 12 stars out of the 13 from the sample  turn out to be long-period spectroscopic binaries with $P > 400$~d, the only exception being HE~1429-0551.

\subsection{Orbital elements}
\label{Sect:orbits}

New\footnote{While the present paper was being refereed, we learned about the orbits obtained by \citet{2015ApJ..NNN..MMMS} for HIP~53522 and HIP~53832, and  by \citet{2015A&A...NNN..MMMH} for HE~0111-1346 and HE~0507-1653, in common with the present sample. The orbital elements for these systems obtained by these independent studies are very similar to ours.} orbital elements for 8 systems are listed in Table~\ref{Tab:orbits} and the corresponding orbital solutions are found in Figs.~\ref{Fig:orbit1} -- \ref{Fig:orbit6} and \ref{Fig:jitter2}--\ref{Fig:jitter3}. The orbits are not fully constrained yet for HD~26 and HD~76396 (in the former case, the period could be even longer than the currently estimated 54~yr; that period is currently the longest known among barium, CH and extrinsic S stars; see the rightmost blue cross in  Fig.~\ref{Fig:e-P} below). We note that the $O-C$ residuals are consistent with the HERMES measurement errors except for HD~26,
HD~76396, and HE~1120-2122, where they are much larger. As seen on Figs.~\ref{Fig:orbit6} and \ref{Fig:jitter3},  these large values 
for HD~26 and HD~76396 
are caused by the dispersion of  the old CORAVEL \citep{1979VA.....23..279B} and \citet{1990ApJ...352..709M} measurements, respectively, the standard deviation of the $O-C$ residuals of the 22 HERMES measurements amounting to 0.126~km~s$^{-1}$ in the case of HD~76396. 
For HE~1120-2122, the large $O-C$ residuals must be considered as a warning that the system may either be triple or be subject to velocity jitter caused 
by, e.g.,  envelope pulsations. This will be discussed further in Sect.~\ref{Sect:short-term}.

\subsection{Systems with velocity variations of low amplitudes and periods $\sim$ 1 yr}
\label{Sect:short-term}

Quite surprisingly, about 1-yr-period variations superimposed on the long-term (Keplerian) variations have been found  for the three systems HE~0017+0055, HE~1120-2122, and HD~76396 (Figs.~\ref{Fig:jitter1}--\ref{Fig:jitter3}). 
For the latter two systems, the orbital $O-C$ residuals are larger than the HERMES spectrograph long-term accuracy of 55~m/s, indicating the presence 
of information in the residuals.  The elements of the Keplerian orbit fitted to these residuals are listed in Table~\ref{Tab:orbits_OC}.
For HE~0017+0055, presented in Fig.~\ref{Fig:jitter1}, nearly 8 cycles of ~1-yr periodic velocity oscillation have been observed on top of the long-term trend. We have found that the latter is best modelled by a quadratic polynomial of the form $Vr = 3.252\times 10^{-7} x^2  -0.0374 x + 993.215$, where $x = JD - 2400000$, while the short-term oscillation is adequately represented by a Keplerian binary orbit with a period of 383~d and an eccentricity of 0.14 (Table~\ref{Tab:orbits_OC}). HE~0017+0055 is the topic of a dedicated paper \citep{2015A&A...XXX..NNNH}, since it  combines radial velocities from HERMES/Mercator and from the Nordic Optical Telescope.  

The associated time scale for the short-term velocity variations is in all three cases close to 1~yr. 
This very fact could raise doubts about the validity of our result (being caused by, e.g., an inaccurate barycentric correction). However, for HE~0017+0055, it was obtained independently on the Nordic Optical Telescope \citep{2015A&A...XXX..NNNH}, which thus gives credit to the result. 
Moreover, the many other systems subject to a radial-velocity monitoring with HERMES \citep{2013EAS....64..163G} do not reveal similarly oscillating $O-C$ residuals, and neither do the radial-velocity standard stars, nor the long-period CEMP star HE~0457-1805 in the present sample (Fig.~\ref{Fig:orbit2}).

The remarkable feature of the Keplerian orbits fitted to the ~1-yr-period variations, as listed in Table~\ref{Tab:orbits_OC}, is the very low value of their velocity amplitude (0.1 to 0.6~\kms), which suggests
three possible interpretations: (i) an orbit with a very low inclination; (ii) a companion of very low mass; (iii) a spurious Keplerian solution, the velocity variations originating instead in the (supposedly pulsating) atmosphere of the giant star. 
These possibilities are discussed in more details in \citet{2015A&A...XXX..NNNH}. Here we only stress that \citet{2003AJ....125..293C} 
showed that  low-metallicity giants, especially in globular clusters and the Magellanic Clouds, systematically exhibit radial-velocity jitter for luminosities near or above the RGB tip (i.e., $M_V \le -1.5$ or $\log g \le 1.3$). These radial-velocity variations mimick Keplerian variations, with semi-amplitudes of the order of  1 -- 1.5~\kms\ and periods of the order of 170 - 190~d. They are thus of  shorter periods and larger amplitudes than those observed in our sample (but the metallicities are different). Moreover, the velocity jitter detected by  \citet{2003AJ....125..293C} appears to vary in phase with photometric variations 
of amplitudes about 0.15~mag. As shown by \citet{2003MNRAS.343L..79K}, these photometric variations are probably linked with the A, B and C period-luminosity sequences first identified by \citet{2000PASA...17...18W} in the Magellanic Clouds.   A final conclusion about the nature of the short-term velocity variations reported here in several CH and CEMP stars must await an accurate photometric monitoring of these systems, as well as a good distance estimate which allows the determination of their position in the Hertzsprung-Russell diagram.  Both diagnostics should be testable with the oncoming Gaia data.

\begin{figure}[h]
\includegraphics[width=9cm]{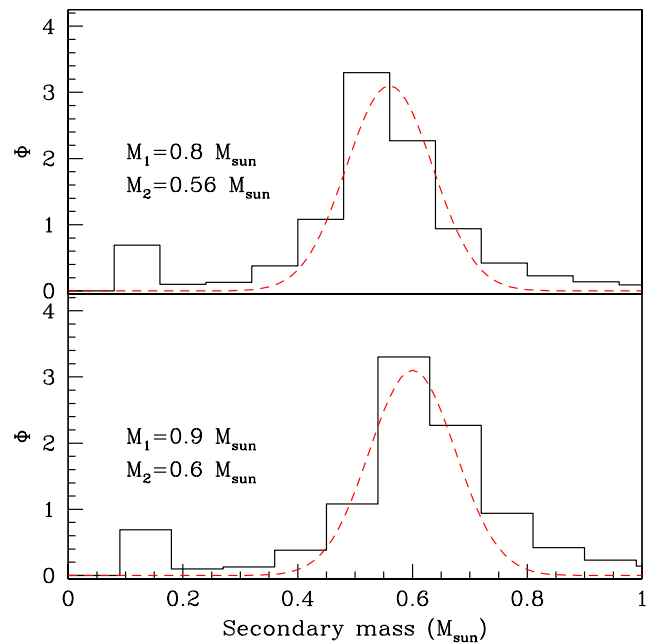}
\caption{\label{Fig:M2}
The distribution of the secondary masses $\Phi(M_2)$ (normalised such that $\int_0^{\infty} \Phi(M_2) \mathrm{d}M_2 = 1$) extracted
from the inversion of the mass functions $f(M_1,M_2)$, assuming a random orientation of the orbital planes on the sky and a fixed primary mass $M_1 = 0.8$~M$_{\odot}$ (top) or $M_1 = 0.9$~M$_{\odot}$ (bottom). For comparison, we also show as the dashed red line a Gaussian distribution of masses centred around $M_2 = 0.56 $~M$_{\odot}$ (top) or $M_1 = 0.6$~M$_{\odot}$ (bottom), in each case with a standard deviation of 0.075~M$_{\odot}$. 
}
\end{figure}

\begin{figure}[h]
\includegraphics[width=9cm]{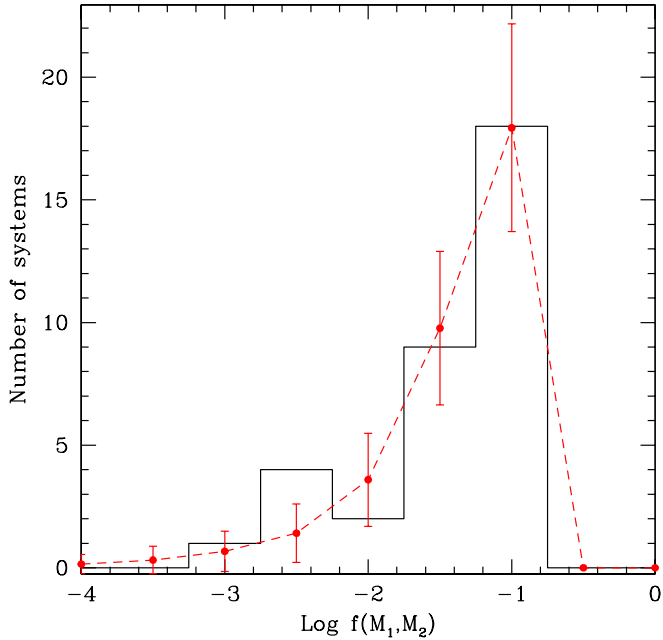}
\caption{\label{Fig:fM}
The observed distribution of  the mass functions $f(M_1,M_2)$ (in solar-mass units; solid black histogram), superimposed on a synthetic distribution (dashed red curve with statistical error bars) resulting from the assumptions of
random orientation of the orbital planes on the sky and fixed value of $Q = 0.094$~M$_{\odot}$. 
}
\end{figure}

\begin{table*}
\caption{\label{Tab:orbits}
Orbital elements for the eight new spectroscopic-binary systems. 
The orbit of HD~26 is not fully constrained yet, its period could be longer than that listed here.
}
\small{
\begin{tabular}{l c c c cccccccccccccc }% neuf colonnes
\hline\hline
Name  &                           HE 0111-1346 & HE 0457-1805 & HE 0507-1653 & HE 1120-2122 & HIP 53522 & HIP 53832  \\
\cline{5-6}
\hline
$P$  (d) &                        $402.74\pm0.05$ & $2724\pm23$ & $404.58\pm0.08$ & $2079.8\pm2.7$ & $361.12\pm0.04$ & $322.84\pm0.08$ \\
$e$  &                              $0.022\pm0.001$ &  $0.191\pm0.007$ & $0.027\pm0.001$ & $0.153\pm0.002$ &  $0.361\pm0.003$ & $0.055\pm0.001$  \\
$\omega$  ($^{\circ}$) &  $271\pm3$       & $64\pm1$ &     $242\pm3$      & $350.5\pm0.9$   &  $70.7\pm0.3$ & $315\pm1$ \\
$V_{0}$ (km/s)&             $38.14\pm0.01$ & $66.80\pm0.03$ & $350.20\pm0.01$      & $108.65\pm0.01$ &  $32.49\pm0.09$ & $-2.28\pm0.01$ \\
$K_1$  (km/s)&                $12.59\pm0.02$  & $6.62\pm0.03$ & $7.18\pm0.01$      & $6.593\pm0.006$ &  $13.22\pm0.02$ &  $11.8\pm0.02$  \\
$a_{1}\;\sin i$ ($10^9$ m) &  $69.7\pm0.1$  & $243\pm3$ & $39.9\pm0.1$    & $186.3\pm0.5$  &  $61.2\pm0.2$ & $52.3\pm0.1$ \\
$f(M)$ (M$_{\odot}$) &   $0.0831\pm0.0003$ &$0.077\pm0.002$ & $0.0155\pm0.0001$ & $0.0596\pm0.0003$  & $0.070\pm0.001$ & $0.0547\pm0.0003$ \\
$T_0$ (-2\,400\,000)&   56\,928$\pm3$ & 56\,180$\pm6$ & $56\,010\pm3$ & $58\,976\pm3$  &56\,862$\pm1$ & 56\,114$\pm1$  \\
$\sigma(O-C)$ (km/s) &    0.11 & 0.04 & 0.14 & 0.42 & 0.07 & 0.04   \\
$N$                            & 59 & 29 & 43 & 40   & 57 & 38  \\
\hline
\end{tabular}
\vspace{5mm}

\begin{tabular}{l c c c cccccccccccccc }% neuf colonnes
\hline\hline
Name  &                           HD 76396 & HD 26   \\
\hline
$P$  (d) &                         $13611\pm109$  & $19634\pm2812$ \\
$e$ &                                $0.41\pm0.02$  & $0.08\pm0.03$ \\
$\omega$  ($^{\circ}$) &   $241\pm7$    & $289\pm48$\\
$V_{0}$ (km/s)&              $-61.7\pm0.2$ & $-214.2\pm0.5$  \\
$K_1$  (km/s)&                 $4.0\pm0.3$  & $3.3\pm0.3$ \\
$a_{1}\;\sin i$ ($10^9$ m) &$689\pm56$ & $899\pm226$ \\
$f(M)$ (M$_{\odot}$) &   $0.07\pm0.02$ & $0.075\pm0.037$  \\
$T_0$ (-2\,400\,000)&      60\,243$\pm234$ & 68\,682$\pm4748$ \\
$\sigma(O-C)$ (km/s) &     0.70 & 0.40 \\
$N$                            &    68 & 110             \\
\hline
\end{tabular}

}

\end{table*}

\begin{table*}
\caption{\label{Tab:orbits_OC}
Orbital elements fitting the $O-C$ residuals of three systems. 
}
\begin{tabular}{l c c c cccccccccccccc }% neuf colonnes
\hline\hline
Name  &                           HE 0017+0055$^a$  & HE 1120-2122 & HD 76396  \\
\cline{5-6}
\hline
$P$  (d) &                        $383\pm1$ &  $363.6\pm1.4$ & $377\pm12$ \\
$e$  &                              $0.14\pm0.03$ &  0.0 &  $0.38\pm0.26$  \\
$\omega$  ($^{\circ}$) &  $212\pm15$       &  -    &   $36\pm77$ \\
$K_1$  (km/s)&                $0.60\pm0.04$  &  $0.9\pm0.3$ &   $0.11\pm0.06$  \\
$a_{1}\;\sin i$ ($10^9$ m) &  $3.1\pm0.2$  &  $4.5\pm1.3$ &  $0.54\pm0.36$ \\
$f(M)$ (M$_{\odot}$) &   $(8.3\pm1.7)\times10^{-6}$ & $(2.7\pm3.1)\times 10^{-5}$ & $(0.4\pm1.5)\times 10^{-7}$ \\
$T_0$ (-2\,400\,000)&      56\,266.7$ \pm 15.7$& 57\,428.9$^b$ & 57\,041$\pm62$  \\
$\sigma(O-C)$ (km/s) &    0.14 &  0.23 & 0.058   \\
$N$                            &    60  & 40    & 37  \\
\hline
\end{tabular}

$^a$ Orbit from \citet{2015A&A...XXX..NNNH}; $^b$ Epoch of maximum velocity.

\end{table*}

\begin{figure}[h]
\includegraphics[width=10cm]{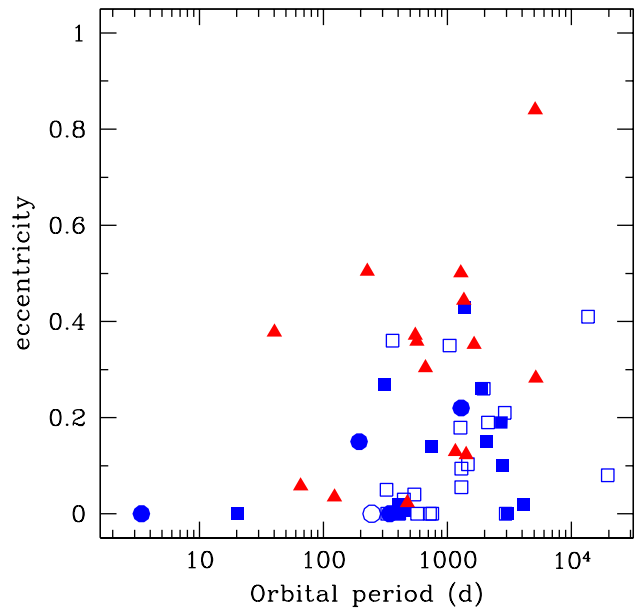}
\caption{\label{Fig:CH_CEMP}
Comparison of CEMP stars (filled symbols) and CH stars (open symbols) in the period --  eccentricity ($P - e$) diagram. Note especially the short-period dwarf-carbon star HE~0024-2523. Dwarf carbon stars are represented by circles, giant carbon stars by squares (CEMP stars with unknown gravities in Table~\ref{Tab:e-P} have been assigned the giant status by default). For comparison, the sample of low-metallicity giants studied by \citet{2003AJ....125..293C} has been represented by red triangles.
}
\end{figure}

\begin{figure}[h]
\includegraphics[width=9.5cm]{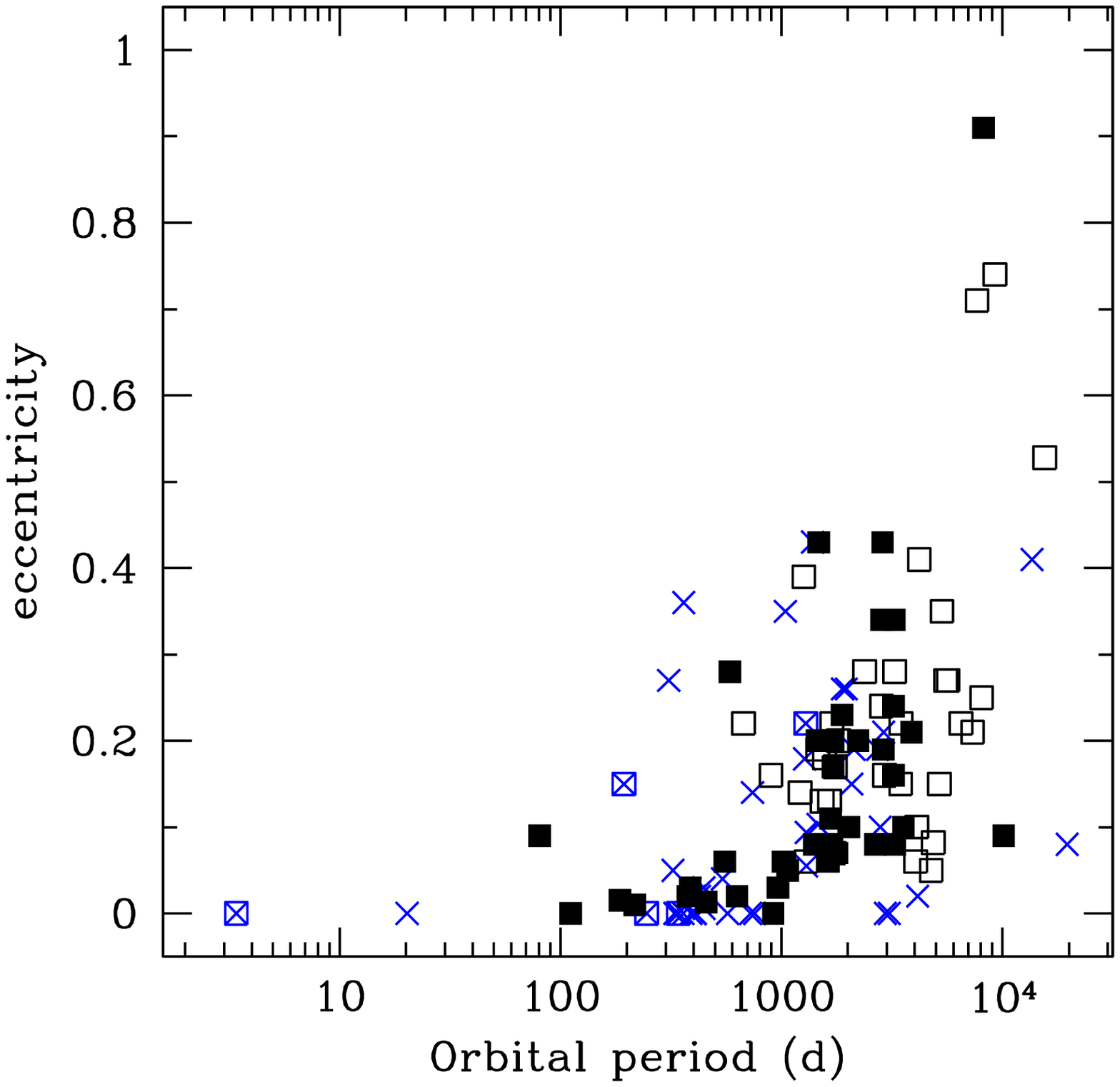}
\caption{\label{Fig:e-P}
The $P - e$ diagram for CEMP-s and CH stars (crosses; squared crosses correspond to carbon dwarfs), and barium stars \protect\citep[from][ filled squares: strong barium stars -- Ba4 and Ba5; open squares: mild barium stars -- Ba1 to Ba3, according to the classification by L\"u et al. 1983]{2013EAS....64..163G}. \protect\nocite{1983ApJS...52..169L}
}
\end{figure}

\section{Analysis}

\subsection{Masses}
\label{Sect:masses}

We perform here the analysis of the mass-function distribution, following the method outlined by
\citet{1992btsf.work...26B}, \citet{1993A&A...271..125B}, \citet{1994InvPr..10..533C}, and \citet{2010A&A...524A..14B,2012ocpd.conf...41B}, with the mass functions collected in Table~\ref{Tab:e-P}. The method gives access to the distribution of the masses of the secondary (denoted $M_2$ in what follows), under the assumption of a random orientation of the orbital planes, and with some {\it a priori} knowledge of the primary masses (denoted $M_1$). In the present situation, we may only rely on their low metallicity to infer that they should have a low mass. The 5 CH-like stars from Sperauskas et al. 2015 in Table~\ref{Tab:e-P} were therefore not included in the mass-function analysis because of their almost solar metallicity. We restrict as well our analysis to the sample of giant stars (from Table~\ref{Tab:e-P}) so as to be able to use the argument that, since they have reached the giant branch, their main-sequence lifetime should be shorter than the age of the Galaxy. We therefore adopt 0.8~--~0.9~M$_{\odot}$ as a typical mass for the primary star [fortunately, the method is not too sensitive to that assumption, given the way $f(M_1,M_2) = M_2^3 \sin^3 i /(M_1+M_2)^2$   depends on $M_1$]. Fig.~\ref{Fig:M2} presents the distribution of  the secondary masses $M_2$, which are peaked in the range 0.5~--~0.7~M$_{\odot}$ (with one exception), as expected for white-dwarf companions originating from the thermally-pulsing asymptotic giant branch \citep{2010ApJ...712..585F}. 
To confirm the robustness of that conclusion, Fig.~\ref{Fig:fM} shows the good agreement between the observed distribution of  the mass functions $f(M_1,M_2)$ (solid black histogram), and a synthetic distribution resulting from the assumptions of
random orientation of the orbital planes on the sky and a fixed value of 0.094~M$_{\odot}$ for $Q = M_2^3 /(M_1+M_2)^2$. Indeed, following \cite{1990ApJ...352..709M}, the distribution of $f(M_1,M_2)$ for the CH systems they studied (ref. 5 in Table~\ref{Tab:e-P}) is compatible with a Gaussian distribution of $Q$ with a mean value of 0.094~M$_{\odot}$ and a standard deviation of 0.013~M$_{\odot}$. At face value, this implies $M_2=0.56$~M$_{\odot}$ for $M_1=0.8$~M$_{\odot}$, and $M_2=0.60$~M$_{\odot}$ for $M_1=0.9$~M$_{\odot}$. This is exactly what is found by our inversion method, applied on a larger sample of CH and CEMP systems.

\begin{figure}[h]
\includegraphics[width=9.5cm]{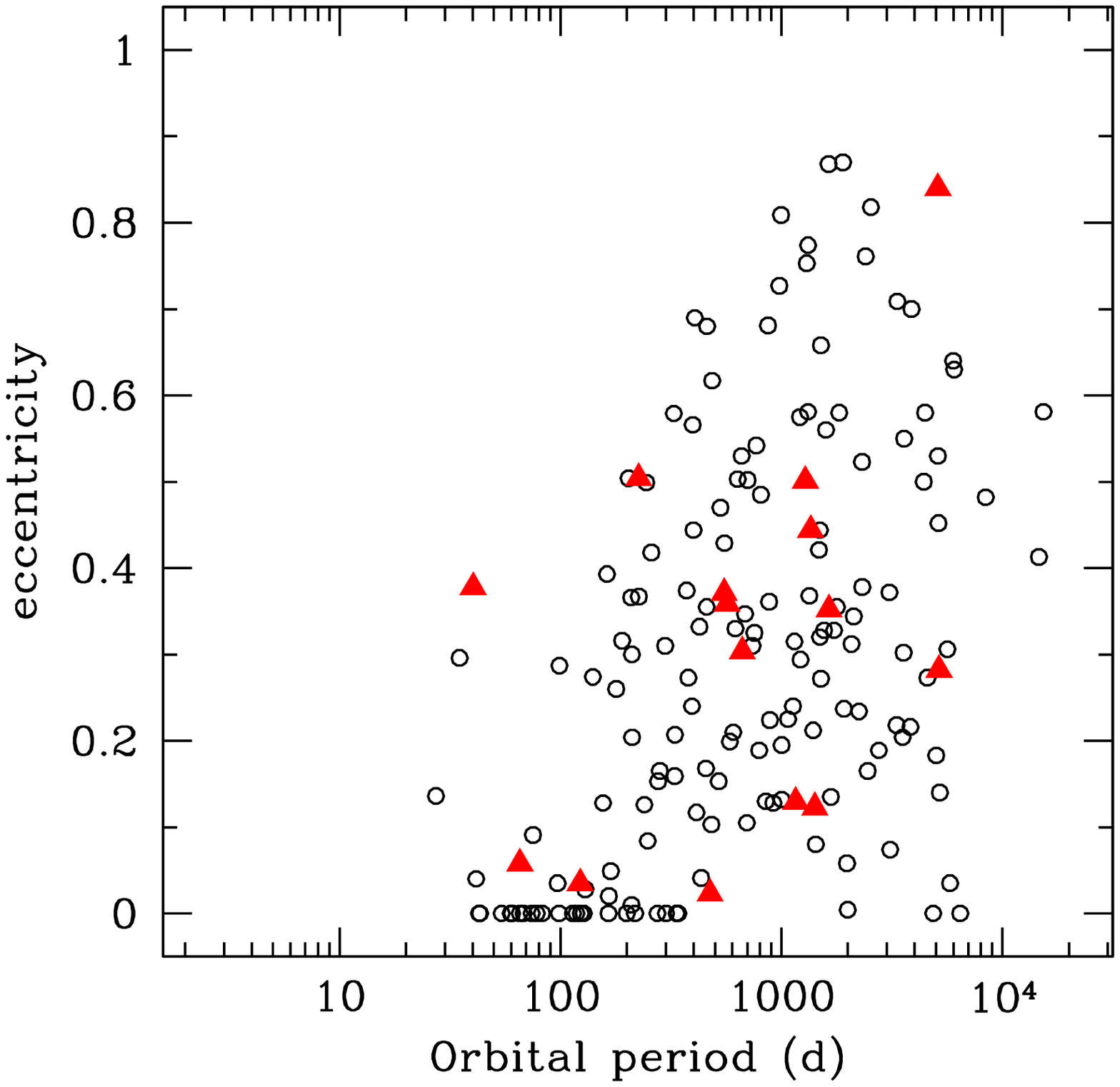}
\caption{\label{Fig:elogP_mdG}
The $P - e$ diagram for normal K giants in open clusters \protect\citep[black open circles, from][]{2007A&A...473..829M} and for non-CEMP low-metallicity giants \citep[red triangles, from][]{2003AJ....125..293C}.
}
\end{figure}

\begin{figure}[h]
\includegraphics[width=9.5cm]{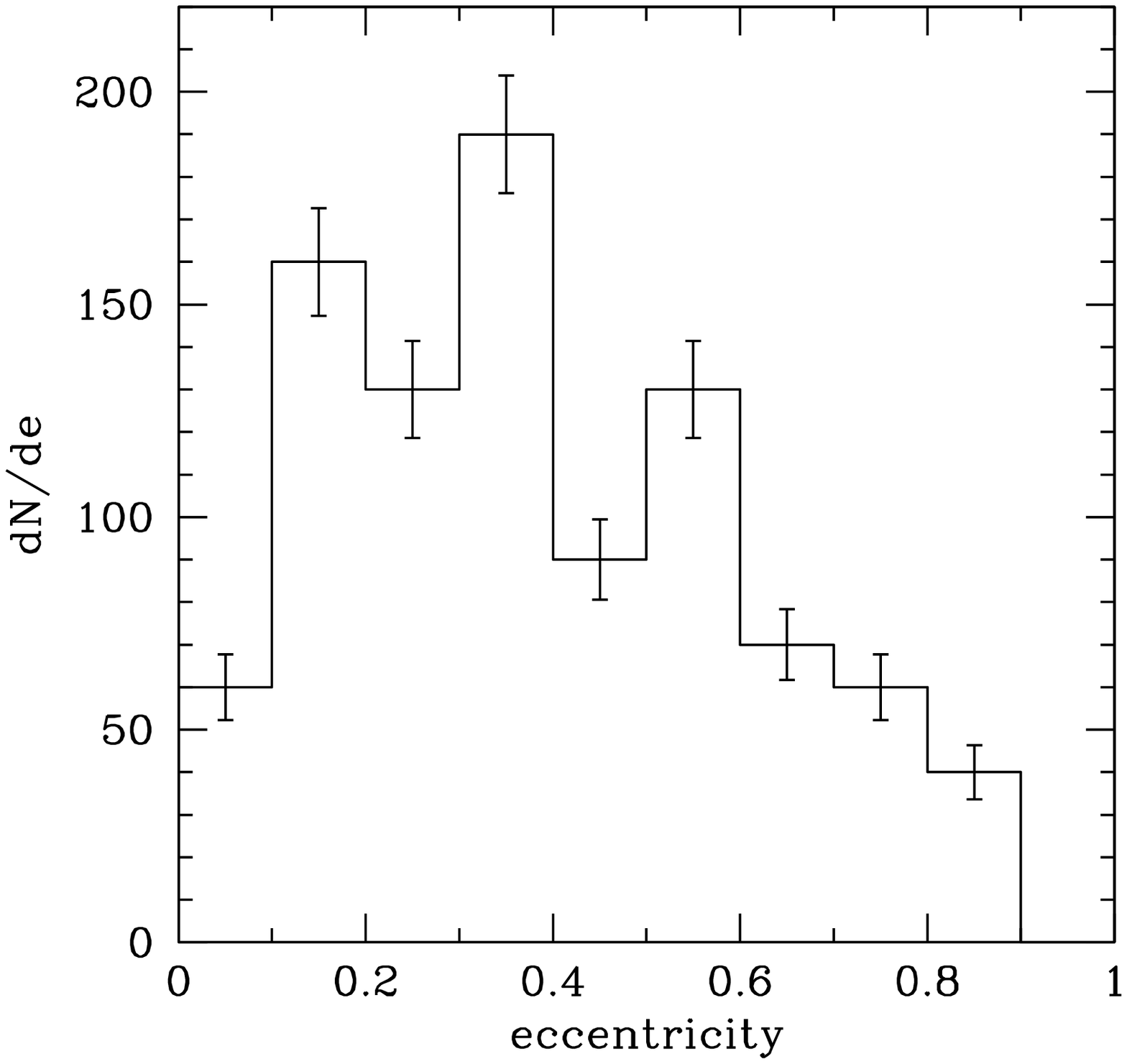}
\caption{\label{Fig:dNde}
The distribution d$N$/d$e$ of the number of systems per eccentricity bins for  normal K giants in open clusters  with periods in excess of 400~d \protect\citep{2007A&A...473..829M}.
}
\end{figure}

\subsection{Period -- eccentricity  diagram}
\subsubsection{CEMP-s vs. CH systems}
\label{Sect:elogP}

We now investigate the period --  eccentricity ($P - e$) diagram for CH and CEMP-(r)s stars, adding  to the 18 systems already available in the literature the 8 new orbits from Table~\ref{Tab:orbits}, the 8 new well-constrained  orbits from \citet{2015A&A...NNN..MMMH} and the 5 new orbits of CH-like stars from \citet{2015ApJ..NNN..MMMS} that we merge with the CH systems.
For the sake of completeness, the data used to draw the $P - e$ diagram of Fig.~\ref{Fig:CH_CEMP} have been collected in Table~\ref{Tab:e-P}, with references indicated. No differences between CEMP-s and CH systems are apparent from Fig.~\ref{Fig:CH_CEMP}, so that in the remainder of this paper, they will be treated as a single group. The orbital similarity between these two groups is a further indication that their different names stem from historical reasons, but there is no physical difference between them, apart from the fact that CH systems always involve giant primaries, whereas 
CEMP systems comprise  as well  dwarf primaries (see the d/g column in Table~\ref{Tab:e-P}). Long-standing historical equivalents of the dwarf CEMP-s stars are the subgiant CH stars \citep{1991ApJS...77..515L} and the formerly known dwarf carbon stars like G77-61 \citep[][ see Table 9.5 of Jorissen 2004 for a review]{1986ApJ...300..314D} \nocite{Jorissen2004} or the ''carbon dwarf wearing a Necklace'', the central star of the planetary nebula G054.2 -- 03.4 \citep[][]{2011MNRAS.410.1349C,2013MNRAS.428L..39M}. The latter has not been listed explicitly in Table~\ref{Tab:e-P}, because it is not formally known as a CH or CEMP star. It has an orbital period of 1.16~d, very similar to that of the CEMP dwarf HE 0024-2523  \citep[3.4~d;][]{2003AJ....125..875L}. 

The orbital-period range spanned by CH/CEMP-s stars is very extended, from a few days for the CEMP dwarf HE~0024-2523 to the (estimated) 54~yr period of the CH star HD~26 (Fig.~\ref{Fig:CH_CEMP}).  Given their smaller size, dwarf carbon stars may fit into systems that are closer (and hence have shorter orbital periods) than those involving giant primaries.  Orbital periods as short as that of HE~0024-2523 are not unusual among systems involving main-sequence primaries  \citep[e.g.,][]{1992btsf.work..155M}. But HE~0024-2523 is supposed to be the post-mass-transfer stage of a system that formerly involved an AGB star. Since the AGB star could not be hosted in a system that close, the current period of 3.4~d must be the outcome of a significant orbital shrinkage. Such a shrinkage is expected in binary-evolution channels involving a common envelope\footnote{The ``carbon dwarf wearing a Necklace'' mentioned above is another example of a short-period (1.16~d) post-common-envelope system \protect\citep{2011MNRAS.410.1349C,2013MNRAS.428L..39M}.}, as it must happen when an AGB star overflows its Roche lobe \citep[e.g.,][]{1995MNRAS.277.1443H,2003ASPC..303..290P}. 
\citet{2015A&A...576A.118A} were able to account for the short orbital period of HE~0024-2523 in such a common-envelope scenario, provided that the envelope ejection was very inefficient (i.e., the fraction of the orbital energy used to unbind the common envelope is small, a few percents only).
This system furthermore raises the (unsolved) question of whether or not accretion is possible during the common-envelope stage \citep{2008ApJ...672L..41R,2015ApJ...803...41M}. In the \citet{2015A&A...576A.118A} scenario for HE~0024-2523, accretion did occur during a brief phase of wind accretion or wind Roche-lobe overflow prior to the common-envelope stage. A similar requirement was invoked by \citet{2013MNRAS.428L..39M} 
for the carbon star at the center of the Necklace planetary nebula.

At the other end of the period range (with periods in excess of $10^4$~d) lie HD~26 and HD~76396 (Table~\ref{Tab:e-P}), and possibly as well HD~145777, HD~187216, HE~1429-0551 and HE~2144-1832 if these would finally turn out to be very long-period binaries. The corresponding frequency of CH/CEMP-s systems with orbital periods in excess of $\sim 10^4$~d
must thus be of the order of 2/44 (=4.5\%), and possibly up to 6/44 (=14\%).  These frequencies may be used to constrain the binary-evolution models of
\citet{2015A&A...581A..22A},  which predict a lot more of such long-period systems (their Fig.~8) than is actually observed. A more detailed comparison is beyond the scope of the present paper, however.

\begin{figure}[h]
\includegraphics[width=9.5cm]{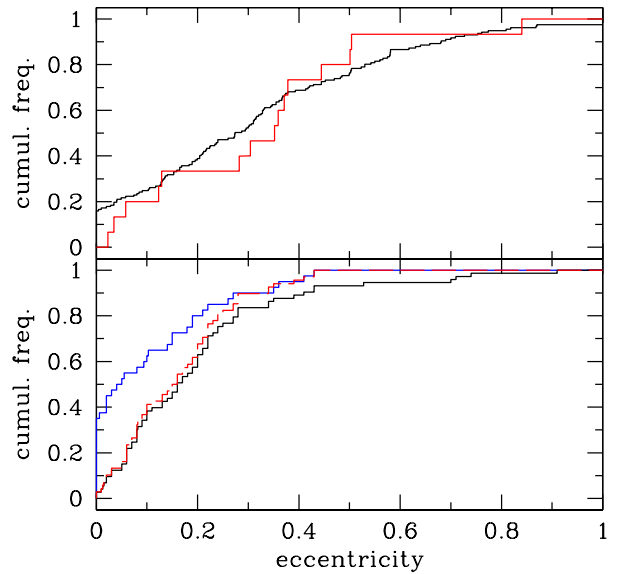}
\caption{\label{Fig:frequency_e}
The cumulative distribution of eccentricities for CH/CEMP-s stars (blue curve in lower panel) and barium stars (black curve in lower panel). The solid black line corresponds to the full barium-star sample, whereas the dashed red line corresponds to the barium-star sample restricted to $e < 0.5$. In the upper panel, the red line corresponds to metal-deficient non-CEMP giants and the solid black curve to K giants in open clusters.
}
\end{figure}

\subsubsection{CH/CEMP-s vs. metal-poor non-CEMP systems}

Fig.~\ref{Fig:CH_CEMP} also compares CH/CEMP-s systems with binary metal-poor giants  that do not belong to the CH/CEMP-s family \citep[from ][]{2003AJ....125..293C}. The latter systems do not bear thus chemical signatures of mass transfer, and may therefore be considered as pre-mass-transfer systems.  
It is very clear that, at a given orbital period,  the post-mass-transfer CH/CEMP-s systems have on average a smaller eccentricity.   A similar difference
between pre- and post-mass-transfer systems has been observed among the solar-metallicity counterparts of CH/CEMP-s systems, namely the barium systems: 
at a given orbital period, barium systems have a smaller average eccentricity than normal K giants \citep{2009A&A...498..489J}, as may be inferred from the comparison of Figs.~\ref{Fig:e-P} and \ref{Fig:elogP_mdG}.

Because the sample of K giants from open clusters mostly (although not exclusively; Van der Swaelmen et al., in preparation) contains pre-mass-transfer 
binaries, it may serve to define the initial distribution of eccentricities, of interest to modellers of binary-population syntheses. It is 
plotted in Fig.~\ref{Fig:dNde} for a sample restricted to $P > 400$~d, because the eccentricity distribution at shorter periods has been altered by circularisation processes.

\subsubsection{CH/CEMP-s vs. barium systems}

In  Fig.~\ref{Fig:e-P}, we compare CH/CEMP-s systems, considered as a single group, with barium systems  \citep[whose orbital elements are taken from][]{2013EAS....64..163G}. 
Overall, the two families cover the same region of the $P - e$ diagram, with nevertheless a few notable differences:
\begin{itemize}
\item There are two CEMP stars with much shorter periods than barium systems, as already discussed in Sect.~\ref{Sect:elogP}.
\item There is a lack of large-eccentricity systems among CH/CEMP-s stars. This effect could result from an observational bias, since 
systems with high eccentricities often require a longer time coverage to derive the corresponding orbits, and CH/CEMP-s were not monitored as long as barium systems were. 
\item The proportion of low-eccentricity systems is much larger among CH/CEMP-s systems. This is clearly apparent on the lower panel of Fig.~\ref{Fig:frequency_e}, which compares the cumulative frequency distributions of eccentricities for barium and CH/CEMP-s stars. The shift between the cumulative distributions at low eccentricities is clearly not a consequence of the difference already reported at large eccentricities, as shown by the red dashed curve in  Fig.~\ref{Fig:frequency_e}, which corresponds to the sample of barium stars restricted to systems with $e < 0.5$.  The difference persists between the CH/CEMP-s sample  and this  sample of barium stars restricted to low eccentricities. The tendency for the low-metallicity CH/CEMP-s systems to have lower eccentricities than their solar-metallicity barium-system counterparts must have some physical origin, yet to be elucidated, related to mass transfer. The role of mass transfer in this eccentricity difference is deduced from the fact that a similar comparison for systems with no chemical anomalies (i.e., non-CEMP low-metallicity giants studied by   \citet{2003AJ....125..293C} and normal K giants in open clusters studied by \citet{2007A&A...473..829M}) reveals no difference in the eccentricity distributions, despite the difference in metallicity (Fig.~\ref{Fig:elogP_mdG} and upper panel of Fig.~\ref{Fig:frequency_e}), when the systems involved are pre-mass-transfer ones. 
\end{itemize}

One  noteworthy similarity between barium and giant CH/CEMP-s systems is the fact that both families seem to separate in two groups, 
1000~d representing the threshold between these: on the one hand, 
the shorter-period systems with mostly circular or almost circular orbits, 
and on the other hand, longer-period systems with no circular orbits (with two exceptions among CEMP systems to be discussed below).  The presence of these two groups in families with so vastly 
different metallicities as CH/CEMP-s and barium stars reinforces the reality of the observed distinction\footnote{The binaries with blue-straggler primaries -- hence they must be post-mass-transfer binaries as are CH/CEMP-s stars -- studied 
by \protect\citet{2005AJ....129..466C} (their Fig.~4)  reveal as well a large number of nearly circular systems with periods shorter than 1000~d.}.
Population syntheses of  binary systems, for both low  \citep{2009A&A...508.1359I}  and solar \citep{2004MmSAI..75..760B,2010A&A...523A..10I} metallicities, indicate that the observed threshold around 1000~d might correspond to 
the transition between the wind-accretion and RLOF channels, an hypothesis supported by the fact that most of the $P < 1000$~d orbits are circular, or nearly so
(Fig.~\ref{Fig:e-P}). Population syntheses do not predict that this period threshold be very sensitive to metallicity \citep[compare Fig.~2 of ][ for barium stars with  Fig.~12 of Izzard et al. 2009 for CEMP stars]{2010A&A...523A..10I}, as observed.
 
The sample of barium systems reveals moreover the existence of a lack of systems at low $e$ and large $P$ (which we will call the low-$e$ large-$P$ gap in the following). 
This gap is especially clearly apparent as well in samples of  pre- and young main-sequence binaries \citep{1992btsf.work..155M}, and indicates that binaries only form as eccentric systems. In young binaries, the gap extends up to $e = 0.15$. At the shortest periods, the gap disappears, filled by tidally-circularised systems, and the period threshold for the gap depends on the age of the systems and the efficiency with which tides act to circularise the orbits. In evolved systems (like the post-mass-transfer barium and CEMP/CH stars; Fig.~\ref{Fig:e-P}), the gap is still present, although to a more limited extent (up to $e = 0.05$), a signature of orbital evolution during mass transfer \citep[see for example the binary population-synthesis models of ][which confirm that there should be very few of such systems]{2010A&A...523A..10I}. For all samples with orbital eccentricities accurately known, this gap is never populated. 
It is thus likely that the eccentricity has not been derived with enough accuracy for the two CH/CEMP-s systems  (HD~30443 and CS~29497-034) falling into the gap in Fig.~\ref{Fig:e-P}. Low-eccentricity orbits need a good many measurements spread over the whole orbital cycle to have the eccentricity accurately determined, and this condition is sometimes difficult to achieve for long-period systems. The latter remark applies to CS~29497-034 \citep{2005A&A...429.1031B}, whereas HD~30443 has a noisy orbital solution whose eccentricity has been fixed at 0.0 \citep{1990ApJ...352..709M}.

\setlength{\tabcolsep}{3pt}

\begin{table*}
\caption{\label{Tab:e-P}
Complete set of $(P,e)$ values currently available for CEMP-(r)s, CH and CH-like stars, along with the mass functions $f(M_1,M_2)$ analysed in Sect.~\ref{Sect:masses}. The column labeled 'd/g' specifies whether the star is a dwarf or a giant. In that column, 'rs' designates a CEMP-rs star. The stars are ordered according to increasing orbital periods. The quantities $N_i$ and $N_{\rm neut}$ are defined in relation with Eq.~\ref{Eq:s}. 
}
\begin{tabular}{l c c c cccccccccccrrrrrrrr}
\hline
Name & $P$ & $e$ & d/g & $f(M_1,M_2)$ & [Fe/H] & \multicolumn{4}{c}{[$i$/Fe]} && \multicolumn{4}{c}{$\left(N_i(\mathrm{s}) - N_i(\mathrm{scaled-solar}) \right)$} & \multicolumn{1}{c}{$N_{\rm neut}$} & Ref.\\
         &        &         &      &                        &            & & & &  && \multicolumn{4}{c}{$\times (A_i - 56)$} & $\times 10^{-4}$ \\
\cline{7-10}\cline{12-15}
         & (d) & & & (M$_{\odot}$) & & Sr & Zr & La & Pb && Sr & Zr & La & Pb\\
\hline
HE 0024-2523  & 3.41 &  0.     &   d &  0.050    & -2.7 &   0.34 & - & 1.8 & 3.3 && 60 & - &  159 & 69443 & 6.97 & 1 \\
HE 1046-1352  & 20.2 &  0.     &  ?  &  0.057   & -3.7  & & & & & &&&&&&19 \\
CS 29509-027  & 194  &  0.15 &  d  & 0.001  & -2.0 & & & & & &&&&&&2 \\
G77-61            &  245.5&  0.   &  d  & -          &  -4.0 & & & & & &&&&&&3,9\\
HE 1523-1155  &  309.3&  0.27& g   & 0.004  & -2.2 & & & & & &&&&&&19,20\\
HIP 53832        & 322.8 & 0.05&  g  & 0.056 & -0.8 & & & & & &&&&&&4,10\\
BD $+42^\circ2173$ &  328.3  & 0. & g & 0.005 & - & & & & & &&&&&&5\\
CS 29497-030  & 342  &  0. &  d  & 0.002  & -2.2 & 1.15 &-  & 1.91 & 3.75 && 2308 & - & 714 & 678838 & 68.19 &2,18 \\
HE 0151-0341 &  359.1 &  0. &   g &  0.067 & -2.5 & & & & & &&&&&&11,19\\
HIP 53522       &  361.1 & 0.36 & g & 0.070 & -0.5 & & & & & &&&&&&4  \\
HE 0854+0151 &  389.9 &  0. &  ?  & 0.086  & -1.8 & & & & & &&&&&&19\\
HE 0111-1346   &  402.7 & 0.02 & g & 0.083 & -1.9 & & & & & &&&&&&4,11 \\
HE 0507-1653   &  405.0 & 0.00 & g & 0.015 & -1.8 & & & & & &&&&&&4,11 \\
HD 209621&         407.4 & 0.     & g & 0.074 & -1.9 & 1.0 & 1.80 & 2.4 & 1.9 && 2880 & 9875 & 4049 & 17232 & 3.12 &5,12\\
CS 22948-27    & 426  &  0.02 &  g (rs) & 0.033  & -2.5 & & & & & &&&&&&6,13 \\
BD $+02^\circ3336$ & 445.9& 0.03 & g & 0.054 & -0.5 & & & & & &&&&&&7,10 \\ 
HE 0507-1430 &  447.0 & 0.01 & g & 0.060 & -2.4  & & & & & &&&&&&19,21 \\
HIP 99725     & 541.1  & 0.04  & g &  0.015 & -0.1 & & & & & &&&&&&22 \\
BD $+08^\circ2654$a& 571.1 &  0.0 & g & 0.044  & - & & & & & &&&&&&5\\
HIP 2529    & 725. &   0.0 &  g &   0.001 & - & & & & & &&&&&&22 \\
HE 0002-1037 &  740.9 & 0.14 &  ? & 0.064 & -3.8 & & & & & &&&&&&19\\
HD 5223  & 755.2 &  0.0 & g & 0.081 & -2.1 & & & & & &&&&&&5,14\\
HIP 43042 &  1042 & 0.35  & g &  0.035 & -0.5 & & & & & &&&&&&22 \\
HD 224959& 1273  & 0.18 & g & 0.092 & -2.1 & & & & & &&&&&&5,15\\
CS 22956-028   & 1290  &  0.22 &  d & 0.076  & -2.1 & & & & &&&&&&&2 \\
HD 198269& 1295 & 0.09 & g & 0.107 & -2.2 & - & 0.4 & 1.6 &2.4 && -  & 121 & 315 & 27549 & 2.80 &5,16\\
HD 202851& 1295 & 0.06 &g & 0.083 &-0.7  & & & & & &&&&&&22 \\
HE 0430-1609 &  1382 & 0.43 & ? & 0.002 & -3.0 & & & & & &&&&&&19\\
HD 201626& 1465  & 0.10 & g & 0.022 & -2.1 & - & 0.9 & 1.9 & 2.6 && - & 697 & 801 & 55049 & 5.65 &5,16\\
HE 2312-0758 &  1890  &   0.26 & ? &  0.014 &  -3.5& & & & & &&&&&&19\\
HE 1120-2122  &  2080 & 0.15 & g & 0.060 & - & & & & & &&&&&&4 \\
HIP 80769 & 2129 &  0.19 & g &  0.089 & - & & & & & &&&&&&22 \\
HE 0457-1805  &  2723 & 0.19 & g & 0.078 & -1.5 & & & & & &&&&&&4,11 \\
CS 22942-019 & 2800 & 0.10 & g & 0.036 & -2.7 & & & & & &&&&&&8,17\\
HD 85066 & 2902 &  0.21 & g &0.113   & - &  & & & & &&&&&&7 \\
HD 30443 & 2954 & 0.0 & g & 0.024 & - & & & & & &&&&&&5\\
HE 0319-0215 &  3078 &   0. & ? & 0.025 &-2.3 & & & & & &&&&&&19\\
CS 29497-034 & 4130 & 0.02 & g (rs) & 0.060 & -2.9 & & & & & &&&&&&6\\  
HD 76396 & 13611 & 0.41 & g & 0.070 & - & & & & & &&&&&&4,5\\ 
HD 26      & 19634 & 0.08 & g & 0.075 & -1.3 & - & 0.9 & 2.3 & 2.0 && - & 4396 & 12791 & 86592 & 10.38 &4,16\\
\hline
\end{tabular}

References: (1) \citet{2003AJ....125..875L} (2) \citet{2003ApJ...592..504S} (3) \citet{1986ApJ...300..314D} (4) This work (5) \citet{1990ApJ...352..709M} (6) \citet{2005A&A...429.1031B} (7) \citet{1997PASP..109..256M} (8) \citet{2001AJ....122.1545P} (9) \citet{2005A&A...434.1117P} (10) \citet{2009A&A...508..909Z} (11) \citet{2011AJ....141..102K} (12) \citet{2010MNRAS.404..253G} (13) \citet{2012A&A...548A..34A} (14) \citet{2006MNRAS.372..343G} (15) \citet{2010A&A...509A..93M} (16) \citet{2003A&A...404..291V} (17) \citet{2000AJ....120.1014P} (18) \citet{2004A&A...413.1073S} (19) \citet{2015A&A...NNN..MMMH} (20) \citet{2007ApJ...655..492A} (21) \citet{2007AJ....133.1193B} (22) \citet{2015ApJ..NNN..MMMS} 
\end{table*}

\setlength{\tabcolsep}{4pt}

To end this comparison between the orbital properties ($P - e$ diagram and mass-function distribution) of CH/CEMP-s and barium stars, 
we may conclude that  a mass-transfer scenario similar to that proposed to account for the chemical peculiarities of barium stars has  been at work in CH/CEMP-s stars as well \citep{1980ApJ...238L..35M,1988A&A...205..155B,2001AJ....122.1545P,2005ApJ...625..825L,
2010A&A...509A..93M,2010A&A...523A..10I,2012MNRAS.422..849B,2013A&A...552A..26A,2014MNRAS.441.1217S}.

\section{A correlation between orbital period and s-process overabundance ?}
\label{Sect:sprocess}

The availability of orbital elements, and especially orbital periods, makes it possible to investigate a possible correlation
between the orbital separation and the s-process overabundance level. Such a correlation (or the absence thereof) provides clues about the mass-transfer process responsible for the pollution of the CEMP/CH star: Roche-lobe overflow, wind Roche-lobe overflow, or wind accretion \citep[see][]{2013A&A...552A..26A,2015A&A...576A.118A,2015A&A...581A..22A}.
A tentative investigation of such a correlation was performed by \citet{1992btsf.work..110J}  for barium 
stars, but using a photometric proxy for the s-process overabundance \citep[see also ][]{1994A&A...291..811B}. A general trend was found  albeit with a very large dispersion.  On the other hand, population synthesis models of \citet{2004MmSAI..75..760B} predict a maximum overabundance level for periods around 1000~d, with the overabundance level going down at both longer and shorter periods.

With the present sample, a similar investigation may be attempted for CH/CEMP-s systems. Several caveats have to be expressed, though. First, it is not the orbital
period, but rather the orbital separation, or even the periastron distance, which is the key dynamical parameter controlling the mass transfer.  For the sample considered, we showed in Sect.~\ref{Sect:masses} that the masses are confined in a narrow range, both for the primaries and the secondaries. Moreover, eccentricities remain moderate, so that orbital periods are expected to be good proxies for the periastron distance.  Second, in low-metallicity stars, the s-process 
operation may produce a substantial amount of Pb \citep[e.g.,][ although 
it does not necessarily do so in a systematic way; see the discussion by De Smedt et al. 2014]{1998ApJ...497..388G,2000A&A...362..599G,2001Natur.412..793V,2003A&A...404..291V}\nocite{2014A&A...563L...5D}. At least for the CEMP/CH stars in Table~\ref{Tab:e-P},
when the Pb abundance is available, it is in fact the dominant contributor to the s-process nuclei, as it will become apparent below. Therefore, the investigation of a possible correlation orbital-period -- s-process overabundance should be restricted to those systems for which Pb abundances are available.  Finally, if as it seems likely, $^{13}$C($\alpha$,n)$^{16}$O is the neutron source \citep{1995ApJ...440L..85S,2015Natur.517..174N}, the number of neutrons produced do not scale with metallicity, since they result from primary seeds [protons and $^{12}$C, leading to $^{12}$C(p,$\gamma$)$^{13}$N($\beta$)$^{13}$C($\alpha$,n)$^{16}$O] \citep{1988MNRAS.234....1C}. Assuming otherwise identical structures and mixing scheme, this means that the number of neutrons made available by the above reaction chain is always the same, irrespective of the metallicity. That number of neutrons is expressed relative to the number of hydrogen (not Fe) atoms, and may  
be extracted from the observed abundance pattern in the following way:
\begin{eqnarray}
\label{Eq:s}
N_{\rm neut} &= &\! \Sigma_i  (A_i - 56) \left[ N_i(\mathrm{star}) - N_i(\mathrm{scaled-solar}) \right] \nonumber\\
            &= &\! \Sigma_i  (A_i - 56) \; [ 10^{\left( [i/\mathrm{Fe}] + [\mathrm{Fe/H}] + \log N_i(\mathrm {sun})\right)} \\
             &  &\; \;\; \;\; \;\; \;\; \; \;\; \;\;\; \; \; \; - 10^{\left( [\mathrm{Fe/H}] + \log N_i(\mathrm{sun})  \right) } ], \nonumber
\end{eqnarray}
where $N_i(\mathrm{star})$ is the number of nuclei of species $i$ per $10^{12}$ hydrogen atoms observed in the star, $A_i$ is the atomic mass number, 
and the sum extends on Zr (or Sr if Zr is not available), La, and Pb.
Solar abundances were adopted from Table~3 of \cite{Lodders2008}, namely $\log N_{\mathrm{Sr}}(\mathrm {sun}) = 2.90$, $\log N_{\mathrm{Zr}}(\mathrm {sun}) = 2.57$, $\log N_{\mathrm{La}}(\mathrm {sun}) = 1.19$, $\log N_{\mathrm{Pb}}(\mathrm {sun}) = 2.06$. The available data are collected in Table~\ref{Tab:e-P}.

The quantity $N_{\rm neut}$ is thus a  measure of the neutron contamination, sensitive to dilution.
The key point here is that the number $N_{\rm neut}$ is derived from the fraction of s-process material which found its way  to the CEMP/CH-star atmosphere.  But this material is only a fraction of the total amount originally produced in the inner shell of the AGB star, which 
has subsequently been diluted three times, first when brought from the AGB interior to its outer atmosphere, then when a fraction of the s-process-enriched AGB wind has been captured by the companion star, and finally when that accreted material has been diluted in the CH/CEMP-star envelope (and the latter dilution thus depends on the evolutionary status of the companion, which may have, or not, a deep convective envelope). 
We want to evaluate the second dilution factor, directly related to the accretion cross section, which is in turn a function of the orbital separation. This dilution factor will cause the estimated number of neutrons to be less than the one originally available in the AGB interior, which we suppose to be a constant, the same for all AGB stars irrespective of their mass or metallicity. Thus, any fluctuation in the estimated neutron number is attributed to the variation of the mass accretion efficiency (assuming that the other two dilutions involved are the same in all cases).  
We must hope that the variations of the latter two dilution factors are smaller than the effect related to the mass-accretion efficiency that we try to measure.

As far as one may judge from column $N_{\rm neut}$ in Table~\ref{Tab:e-P}, there is no obvious trend emerging with orbital period.  Therefore, our data confirm once more that the orbital period is not the only parameter governing the level of s-process overabundances in post-mass-transfer systems \citep[see also][ for a recent analysis]{2015A&A...576A.118A}. The dilution factors discussed above (and neglected in the analysis) may play an important role,  or as \citet{2015A&A...XXXA..YYM} have suggested, the mass of the companion WD (a proxy for the luminosity reached by its progenitor on the AGB) could be another important parameter controlling the level of s-process overabundance.

\section{Conclusions}
\label{Sect:conclusions}

We report 8 new   orbits for CH and CEMP-s systems
obtained by combining new radial-velocity measurements from the
HERMES/Mercator spectrograph with old measurements from
\citet{1990ApJ...352..709M}. HIP~53522 not only is a newly discovered binary, 
but is as well a new member of the CH family, as revealed by a detailed abundance analysis. 
Adding these new orbits to the 33 CH/CEMP-s orbits already
available in the literature, their  $P - e$ and $f(M)$ distributions 
have been investigated and reveal no difference between CH and CEMP-s systems. These families should thus not be
treated separately, since their different names stem from historical reasons but cover the same underlying stellar family.
The analysis
of the mass-function distribution of CH and CEMP-s systems points at
companions having a mass peaked in the range 0.5~--~0.7~M$_{\odot}$,
thus supporting the usual scenario of s-process pollution from a former
AGB star, now the white-dwarf companion. The signature of the mass-transfer scenario is also apparent from the comparison between the 
$P - e$ distributions of CH/CEMP-s stars and of non-carbon-enriched low-metallicity giants (thus being likely pre-mass-transfer systems), 
the former having a smaller average eccentricity 
at a given orbital period. 

CH/CEMP-s orbits have then be 
compared to those of barium stars, which are
their higher-metallicity analogs. Despite minor quantitative differences 
(i.e., a larger fraction of low-eccentricity orbits among CH/CEMP-s stars, and a deficit of large-eccentricity systems),
the global distribution of these families in the $P - e$ diagram is similar,
suggesting that they followed the same
binary-evolution channel. It is not clear at this stage whether the different eccentricity distributions result from observational selection biases,
or from the influence of 
metallicity on the outcome of the mass transfer. 

At least three CH/CEMP-s stars (HE~0017+055, HE~1120-2122, and HD~76396) exhibit short-term radial-velocity variations (with periods close to 1~yr), and 
it is not clear yet whether this behaviour is attributable to some intrinsic phenomenon causing velocity jitter, or whether it is due to Keplerian variations of small amplitude \citep[see][ for a discussion of this alternative]{2003AJ....125..293C,2015A&A...XXX..NNNH}.

\begin{acknowledgement}
This research has been funded by the Belgian Science Policy Office under contract BR/143/A2/STARLAB.
T.M. is supported by the FNRS-F.R.S. as temporary post-doctoral researcher  under grant No. T.0198.13.
SvE is FNRS research associate.
Based on observations obtained with the
HERMES spectrograph, supported by the Fund for Scientific Research of Flanders (FWO),
the Research Council of K.U.Leuven, the Fonds National de la Recherche Scientifique
(F.R.S.-FNRS), Belgium, the Royal Observatory of Belgium, the Observatoire de Gen\`eve,
Switzerland and the Th\"uringer Landessternwarte Tautenburg, Germany. 
\end{acknowledgement}

\clearpage

\appendix
\section{Abundance analysis of HIP 53522}
\label{Sect:appendix}

This appendix describes the abundance analysis performed for HIP~53522, since none was available so far\footnote{While the present paper was being refereed, we learned about the abundance analysis of that same star made independently by \citet{2015ApJ..NNN..MMMS}, with results very similar to ours.} for that star first classified as R by  
\citet{1958AJ.....63..477V} and then reclassified as CH by  \citet{1985AJ.....90.2244H} and CH-like by \citet{2015ApJ..NNN..MMMS}. 
The determination of its atmospheric parameters and s-process element abundances follows the same steps as in  \citet{2014A&A...567A..30M}. The HERMES/Mercator spectrum selected for the analysis was obtained on April 10, 2010 (HJD~2455296.567), with an exposure time of 2600~s and a signal-to-noise ratio of about 50. 
%The barycentric velocity correction at midpoint is $-18.931$~\kms. 
The synthetic spectra were convolved with a Gaussian function with full width at half maximum of 6~\kms.

\subsection{Atmospheric parameters}
\label{Sect:atmosphere}
Previous studies of HIP 53522 \citep{2001A&A...369..178B,2012MNRAS.427..343M} have obtained effective temperatures of 4955 and 4761~K,   respectively. Surface gravity and metallicity are determined iteratively using the \bacchus\ pipeline developed by one of the author (TMa) \citep{2014A&A...564A.133J} in the context of the Gaia-ESO survey \citep[][]{2012Msngr.147...25G}. This pipeline is based on the 1D LTE spectrum-synthesis code Turbospectrum \citep{1998A&A...330.1109A,2012ascl.soft05004P}, and allows an automated determination of effective temperature \Teff, surface gravity $\log{g}$, metallicity \FeH\ and microturbulent velocity $\xi$. We use MARCS model atmospheres \citep{2008A&A...486..951G} along with a selection of neutral and singly ionized Fe lines that were selected for the analysis of stellar spectra in the framework of the Gaia-ESO survey. Oscillator strengths are from the VALD database \citep{2000BaltA...9..590K}. Only lines having reduced equivalent widths ($W/\lambda$) lower than 0.03~m\AA/\AA\ are kept in the analysis. We thus obtain the following atmospheric parameters: \Teff\ $=4505\pm160$~K, $\log{g}=1.7\pm0.9$, \FeH\ $=-0.46\pm0.07$ and $\xi=1.5\pm0.1$~\kms. Solar abundances are from \citet{2007SSRv..130..105G}, where $A_\odot(\mathrm{Fe}) = 7.45$.

\begin{figure}
\includegraphics[width=\linewidth]{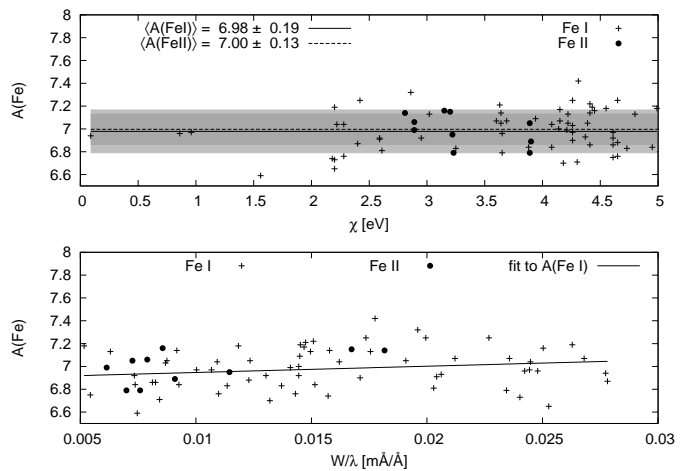}
\caption{\ion{Fe}{i} and \ion{Fe}{ii} abundances (represented by crosses and filled circles, respectively)  derived from individual lines 
as a function of lower  excitation potential $\chi$ (upper panel) and of reduced equivalent width $W/\lambda$ (lower panel). The grey areas represent the standard deviations around the mean abundances of \ion{Fe}{i} and \ion{Fe}{ii}.}
\label{fig:pabu}
\end{figure}

\subsection{s-process element abundances}
\label{Sect:abundances}
%We {\bf derive} abundances of the s-process elements from the first and second peaks using the \bacchus\ pipeline. 
The atomic lines used to derive abundances of s-process elements are listed in Table~\ref{tab:ll}, along with their oscillator strengths, taken from the VALD database \citep{2000BaltA...9..590K}.
% \citet{2014A&A...567A..30M}. 
The list includes the isotopic shifts and hyperfine structure for \ion{Ba}{ii} lines  and the hyperfine structure for \ion{La}{ii}  and \ion{Ce}{ii} lines from \citet{2006PhD}. 
Abundances are derived from the best match with a synthetic spectrum. For barium, the solar-system isotopic mix is adopted.
%Line fitting is essentially based on a least-square minimization method and all lines are visually inspected to check for possible bad fits (due to, e.g., line blends, cosmic hits,...). 
The results of the abundance analysis are presented in Table~\ref{tab:abu}.

An average light-s-process abundance  of [ls/Fe$]=1.20\pm0.10$ is obtained, based on Sr, Y, and Zr abundances, as compared to [hs/Fe$]=1.15\pm 0.10$,  based on Ba, La, and Ce abundances. Thus HIP~53522 is clearly enriched in s-process elements.

\begin{table}
 \caption{Results of the chemical abundance analysis. $A({\rm X})$ is the abundance of species X in the logarithmic scale where $\log A({\rm H}) = 12$.
$\sigma_\mathrm{stat}$ is the line-to-line abundance dispersion. $N$ is the number of lines used for the corresponding species. %In the column labelled 'comments', wavelengths are expressed in nm.
}
%\tiny
 \begin{tabular}{lcrrcrl}
  X & A(X) & [X/H] & [X/Fe] & $\sigma_\mathrm{stat}/\sqrt{N}$ &  $N$ \\ %& comments\\
  \hline \\
  Fe  & 6.99 &$-0.46$ &        & 0.03 & 64 & \\
  \\
  Sr  & 3.66 & $0.74$ & $1.20$ & 0.01 &  2 & \\
  Y   & 2.78 & $0.57$ & $1.03$ & 0.04 & 11 & \\
  Zr  & 3.49 & $0.91$ & $1.37$ & 0.09 &  9 & \\
  \\
  Ba  & 3.23 & $1.06$ & $1.52$ & 0.09 &  4 & \\
  La  & 1.73 & $0.60$ & $1.06$ & 0.03 &  9 & \\
  Ce  & 2.12 & $0.42$ & $0.88$ & 0.04 & 12 & \\
  \hline
 \end{tabular}
 %\normalsize
  \label{tab:abu}
\end{table}
 
\onecolumn
\begin{longtab}
 \begin{longtable}{lccrl}
 \caption{Lines used for deriving the s-process element abundances. $\chi_{\rm exc}$ is the lower excitation potential of the transition. 
A vertical bar to the left of the wavelength values groups hyperfine structure and isotopic shifts for a given line.
\label{tab:ll}}\\
Element& $\lambda$ & $\chi_{\rm exc}$ & \multicolumn{1}{c}{$\log gf$} \\
     & (nm)         & (eV) &\\ 
 \hline
 \endfirsthead
 \caption{Continued.}\\
 Element & $\lambda$ & $\chi_{\rm exc}$ &  \multicolumn{1}{c}{$\log gf$} \\
     & (nm)         & (eV) &\\ 
 \hline
 \endhead
 \hline
 \endfoot

\hline\\
  Sr I & 460.7327 & 0.000 & $-0.570$ \\
  Sr I & 707.0070 & 1.847 & $-0.030$ \\
  Y I  & 552.7547 & 1.398 & $ 0.471$ \\
  Y I  & 643.5004 & 0.066 & $-0.820$ \\
  Y II & 488.3684 & 1.084 & $ 0.265$ \\
  Y II & 490.0120 & 1.033 & $ 0.103$ \\
  Y II & 512.3211 & 0.992 & $-1.219$ \\
  Y II & 528.9815 & 1.033 & $-1.850$ \\
  Y II & 540.2774 & 1.839 & $-0.630$ \\
  Y II & 554.4611 & 1.738 & $-1.090$ \\
  Y II & 554.6009 & 1.748 & $-0.754$ \\ 
  Y II & 572.8890 & 1.839 & $-1.120$ \\
  Y II & 679.5414 & 1.738 & $-1.030$ \\ 
  Zr I & 538.5151 & 0.519 & $-0.710$ \\
  Zr I & 568.0920 & 0.543 & $-1.700$ \\
  Zr I & 573.5690 & 0.000 & $-2.240$ \\
  Zr I & 612.7475 & 0.154 & $-1.060$ \\
  Zr I & 613.4585 & 0.000 & $-1.280$ \\
  Zr I & 733.6066 & 0.519 & $-2.170$ \\
  Zr I & 743.9889 & 0.543 & $-1.810$ \\
  Zr I & 805.8107 & 0.623 & $-2.020$ \\
  Zr I & 807.0115 & 0.730 & $-0.790$ \\
$^{137}$Ba II & \multicolumn{1}{|l}{455.3998} & 0.000 & $-0.666$ \\ 
$^{137}$        & \multicolumn{1}{|l}{455.3999} & 0.000 & $-0.666$ \\ 
$^{137}$      & \multicolumn{1}{|l}{455.4000} & 0.000 & $-1.064$ \\ 
$^{135}$      & \multicolumn{1}{|l}{455.4001} & 0.000 & $-0.666$ \\ 
$^{135}$      & \multicolumn{1}{|l}{455.4002} & 0.000 & $-1.064$ \\ 
$^{135}$      & \multicolumn{1}{|l}{455.4002} & 0.000 & $-0.666$ \\ 
$^{130}$      & \multicolumn{1}{|l}{455.4031} & 0.000 & $ 0.140$ \\ 
$^{132}$      & \multicolumn{1}{|l}{455.4031} & 0.000 & $ 0.140$ \\ 
$^{134}$      & \multicolumn{1}{|l}{455.4031} & 0.000 & $ 0.140$ \\ 
$^{136}$      & \multicolumn{1}{|l}{455.4032} & 0.000 & $ 0.140$ \\ 
$^{138}$      & \multicolumn{1}{|l}{455.4033} & 0.000 & $ 0.140$ \\ 
$^{135}$      & \multicolumn{1}{|l}{455.4048} & 0.000 & $-0.219$ \\ 
$^{135}$      & \multicolumn{1}{|l}{455.4050} & 0.000 & $-0.666$ \\ 
$^{137}$      & \multicolumn{1}{|l}{455.4051} & 0.000 & $-0.219$ \\ 
$^{135}$      & \multicolumn{1}{|l}{455.4052} & 0.000 & $-1.365$ \\ 
$^{137}$      & \multicolumn{1}{|l}{455.4054} & 0.000 & $-0.666$ \\ 
$^{137}$      & \multicolumn{1}{|l}{455.4055} & 0.000 & $-1.365$ 
\medskip\\ 
$^{137}$Ba II & \multicolumn{1}{|l}{493.4030} & 0.000 & $-0.662$ \\ 
$^{135}$      & \multicolumn{1}{|l}{493.4032} & 0.000 & $-0.662$ \\ 
$^{135}$      & \multicolumn{1}{|l}{493.4042} & 0.000 & $-1.361$ \\ 
$^{137}$      & \multicolumn{1}{|l}{493.4042} & 0.000 & $-1.361$ \\
$^{130}$      & \multicolumn{1}{|l}{493.4074} & 0.000 & $-0.157$ \\ 
$^{132}$      & \multicolumn{1}{|l}{493.4074} & 0.000 & $-0.157$ \\ 
$^{134}$      & \multicolumn{1}{|l}{493.4074} & 0.000 & $-0.157$ \\
$^{136}$      & \multicolumn{1}{|l}{493.4075} & 0.000 & $-0.157$ \\ 
$^{138}$      & \multicolumn{1}{|l}{493.4077} & 0.000 & $-0.157$ \\
$^{135}$      & \multicolumn{1}{|l}{493.4091} & 0.000 & $-0.662$ \\ 
$^{137}$      & \multicolumn{1}{|l}{493.4095} & 0.000 & $-0.662$ \\ 
$^{135}$      & \multicolumn{1}{|l}{493.4102} & 0.000 & $-0.662$ \\   
$^{137}$      & \multicolumn{1}{|l}{493.4107} & 0.000 & $-0.662$ 
\medskip\\
$^{135}$Ba II &\multicolumn{1}{|l}{614.1708} & 0.704 & $-0.456$ \\ 
$^{135}$      & \multicolumn{1}{|l}{614.1708} & 0.704 & $-1.264$ \\  
$^{135}$      & \multicolumn{1}{|l}{614.1709} & 0.704 & $-2.410$ \\  
$^{137}$      & \multicolumn{1}{|l}{614.1709} & 0.704 & $-1.264$ \\  
$^{137}$      & \multicolumn{1}{|l}{614.1709} & 0.704 & $-0.456$ \\  
$^{137}$      & \multicolumn{1}{|l}{614.1710} & 0.704 & $-2.410$ \\  
$^{130}$      & \multicolumn{1}{|l}{614.1711} & 0.704 & $-0.030$ \\ 
$^{132}$      & \multicolumn{1}{|l}{614.1711} & 0.704 & $-0.030$ \\ 
$^{134}$      & \multicolumn{1}{|l}{614.1711} & 0.704 & $-0.030$ \\ 
$^{136}$      & \multicolumn{1}{|l}{614.1712} & 0.704 & $-0.030$ \\  
$^{135}$      & \multicolumn{1}{|l}{614.1713} & 0.704 & $-0.662$ \\  
$^{138}$      & \multicolumn{1}{|l}{614.1713} & 0.704 & $-0.030$ \\ 
$^{135}$      & \multicolumn{1}{|l}{614.1714} & 0.704 & $-1.167$ \\  
$^{135}$      & \multicolumn{1}{|l}{614.1715} & 0.704 & $-2.234$ \\  
$^{137}$      & \multicolumn{1}{|l}{614.1715} & 0.704 & $-0.662$ \\  
$^{135}$      & \multicolumn{1}{|l}{614.1716} & 0.704 & $-0.912$ \\  
$^{137}$      & \multicolumn{1}{|l}{614.1716} & 0.704 & $-1.167$ \\  
$^{135}$      & \multicolumn{1}{|l}{614.1717} & 0.704 & $-1.234$ \\  
$^{135}$      & \multicolumn{1}{|l}{614.1717} & 0.704 & $-1.280$ \\  
$^{137}$      & \multicolumn{1}{|l}{614.1717} & 0.704 & $-2.234$ \\  
$^{137}$      & \multicolumn{1}{|l}{614.1718} & 0.704 & $-0.912$ \\  
$^{137}$      & \multicolumn{1}{|l}{614.1719} & 0.704 & $-1.234$ \\  
$^{137}$      & \multicolumn{1}{|l}{614.1719} & 0.704 & $-1.280$ 
\medskip\\ 
$^{135}$Ba II & \multicolumn{1}{|l}{649.6883} & 0.604 & $-1.911$ \\ 
$^{137}$      & \multicolumn{1}{|l}{649.6883} & 0.604 & $-1.911$ \\ 
$^{135}$      & \multicolumn{1}{|l}{649.6888} & 0.604 & $-1.212$ \\ 
$^{137}$      & \multicolumn{1}{|l}{649.6888} & 0.604 & $-1.212$ \\ 
$^{130}$      & \multicolumn{1}{|l}{649.6895} & 0.604 & $-0.406$ \\
$^{132}$      & \multicolumn{1}{|l}{649.6895} & 0.604 & $-0.406$ \\
$^{134}$      & \multicolumn{1}{|l}{649.6895} & 0.604 & $-0.406$ \\
$^{135}$      & \multicolumn{1}{|l}{649.6895} & 0.604 & $-0.765$ \\
$^{137}$      & \multicolumn{1}{|l}{649.6896}& 0.604 & $-0.765$ \\
$^{136}$      & \multicolumn{1}{|l}{649.6897} & 0.604 & $-0.406$ \\
$^{138}$      & \multicolumn{1}{|l}{649.6898} & 0.604 & $-0.406$ \\
$^{135}$      & \multicolumn{1}{|l}{649.6900} & 0.604 & $-1.610$ \\ 
$^{135}$      & \multicolumn{1}{|l}{649.6902} & 0.604 & $-1.212$ \\ 
$^{137}$      & \multicolumn{1}{|l}{649.6902} & 0.604 & $-1.610$ \\ 
$^{137}$      & \multicolumn{1}{|l}{649.6904} & 0.604 & $-1.212$ \\ 
$^{135}$      & \multicolumn{1}{|l}{649.6906} & 0.604 & $-1.212$ \\ 
$^{137}$      & \multicolumn{1}{|l}{649.6909} & 0.604 & $-1.212$ 
\medskip\\
La II & \multicolumn{1}{|l}{466.2478} & 0.000 & $-2.952$ \\     
      & \multicolumn{1}{|l}{466.2482} & 0.000 & $-2.511$ \\    
     & \multicolumn{1}{|l}{466.2486} & 0.000 & $-2.240$ \\    
       & \multicolumn{1}{|l}{466.2491} & 0.000 & $-2.253$ \\     
       & \multicolumn{1}{|l}{466.2492} & 0.000 & $-2.137$ \\    
       & \multicolumn{1}{|l}{466.2493} & 0.000 & $-2.256$ \\
       & \multicolumn{1}{|l}{466.2503} & 0.000 & $-2.511$ \\    
       & \multicolumn{1}{|l}{466.2505} & 0.000 & $-2.056$ \\    
      & \multicolumn{1}{|l}{466.2507} & 0.000 & $-1.763$ \\ 
La II & 474.8726 & 0.927 & $-0.540$ \\
La II & 529.0818 & 0.000 & $-1.650$ \\
La II & \multicolumn{1}{|l}{530.1908} & 0.403 & $-3.065$ \\     
      & \multicolumn{1}{|l}{530.1913} & 0.403 & $-2.266$ \\    
      & \multicolumn{1}{|l}{530.1917} & 0.403 & $-2.391$ \\    
      & \multicolumn{1}{|l}{530.1946} & 0.403 & $-3.483$ \\     
      & \multicolumn{1}{|l}{530.1953} & 0.403 & $-2.300$ \\     
      & \multicolumn{1}{|l}{530.1958} & 0.403 & $-2.120$ \\ 
      & \multicolumn{1}{|l}{530.2001} & 0.403 & $-2.483$ \\     
      & \multicolumn{1}{|l}{530.2008} & 0.403 & $-1.913$ \\     
      & \multicolumn{1}{|l}{530.2067} & 0.403 & $-1.742$ 
      \medskip\\

La II& \multicolumn{1}{|l}{530.3513} & 0.321 & $-1.874$ \\
      & \multicolumn{1}{|l}{530.3513} & 0.321 & $-2.363$ \\
      & \multicolumn{1}{|l}{530.3514} & 0.321 & $-3.062$ \\
      & \multicolumn{1}{|l}{530.3531} & 0.321 & $-2.167$ \\
      & \multicolumn{1}{|l}{530.3532} & 0.321 & $-2.247$ \\
      & \multicolumn{1}{|l}{530.3532} & 0.321 & $-2.622$ \\
      & \multicolumn{1}{|l}{530.3546} & 0.321 & $-2.366$ \\
      & \multicolumn{1}{|l}{530.3546} & 0.321 & $-2.622$ \\
      & \multicolumn{1}{|l}{530.3547} & 0.321 & $-2.351$ \\
La II & 580.5773 & 0.126 & $-1.560$ \\
La II & 593.6210 & 0.173 & $-2.070$ \\
La II & \multicolumn{1}{|l}{626.2113} & 0.403 & $-3.047$ \\    
      & \multicolumn{1}{|l}{626.2114} & 0.403 & $-2.901$ \\    
      & \multicolumn{1}{|l}{626.2132} & 0.403 & $-2.705$ \\    
      & \multicolumn{1}{|l}{626.2134} & 0.403 & $-2.718$ \\    
      & \multicolumn{1}{|l}{626.2135} & 0.403 & $-3.378$ \\    
      & \multicolumn{1}{|l}{626.2164} & 0.403 & $-2.471$ \\    
      & \multicolumn{1}{|l}{626.2166} & 0.403 & $-2.596$ \\    
      & \multicolumn{1}{|l}{626.2169} & 0.403 & $-3.269$ \\
      & \multicolumn{1}{|l}{626.2208} & 0.403 & $-2.286$ \\     
      & \multicolumn{1}{|l}{626.2212} & 0.403 & $-2.535$ \\    
      & \multicolumn{1}{|l}{626.2215} & 0.403 & $-3.290$ \\    
      & \multicolumn{1}{|l}{626.2266} & 0.403 & $-2.130$ \\     
      & \multicolumn{1}{|l}{626.2271} & 0.403 & $-2.531$ \\     
      & \multicolumn{1}{|l}{626.2275} & 0.403 & $-3.400$ \\
      & \multicolumn{1}{|l}{626.2338} & 0.403 & $-1.994$ \\     
      & \multicolumn{1}{|l}{626.2343} & 0.403 & $-2.597$ \\     
      & \multicolumn{1}{|l}{626.2348} & 0.403 & $-3.612$ \\
      & \multicolumn{1}{|l}{626.2422} & 0.403 & $-1.873$ \\     
      & \multicolumn{1}{|l}{626.2429} & 0.403 & $-2.802$ \\     
      & \multicolumn{1}{|l}{626.2434} & 0.403 & $-4.015$ 
\medskip\\ 
La II & \multicolumn{1}{|l}{639.0455} & 0.321 & $-2.012$ & \\ 
      & \multicolumn{1}{|l}{639.0468} & 0.321 & $-2.183$ & \\ 
      & \multicolumn{1}{|l}{639.0468} & 0.321 & $-2.752$ & \\ 
      & \multicolumn{1}{|l}{639.0479} & 0.321 & $-2.570$ & \\ 
      & \multicolumn{1}{|l}{639.0479} & 0.321 & $-3.752$ & \\ 
      & \multicolumn{1}{|l}{639.0480} & 0.321 & $-2.390$ & \\ 
      & \multicolumn{1}{|l}{639.0489} & 0.321 & $-2.536$ & \\
      & \multicolumn{1}{|l}{639.0489} & 0.321 & $-3.334$ & \\
      & \multicolumn{1}{|l}{639.0490} & 0.321 & $-2.661$ & \\
      & \multicolumn{1}{|l}{639.0496} & 0.321 & $-3.100$ & \\
      & \multicolumn{1}{|l}{639.0497} & 0.321 & $-2.595$ & \\
      & \multicolumn{1}{|l}{639.0498} & 0.321 & $-3.079$ & \\
      & \multicolumn{1}{|l}{639.0502} & 0.321 & $-2.954$ & \\
      & \multicolumn{1}{|l}{639.0503} & 0.321 & $-2.778$ & \\
      & \multicolumn{1}{|l}{639.0506} & 0.321 & $-2.857$ & 
\medskip\\
Ce II & \multicolumn{1}{|l}{434.9768} & 0.529 & $-0.520$ & \\  
     & \multicolumn{1}{|l}{434.9789} & 0.701 & $-0.350$ & \\
Ce II & 456.2359 & 0.478 & $ 0.230$ & \\ 
Ce II & \multicolumn{1}{|l}{462.8161} & 0.516 & $ 0.200$ & \\   
      & \multicolumn{1}{|l}{462.8185} & 1.194 & $-3.280$ & \\  
      & \multicolumn{1}{|l}{462.8239} & 1.366 & $-0.430$ & \\
Ce II & 477.3941 & 0.924 & $-0.390$ & \\
Ce II & \multicolumn{1}{|l}{518.7503} & 0.559 & $-3.330$ & \\  
      & \multicolumn{1}{|l}{518.7458} & 1.212 & $ 0.150$ & \\  
      & \multicolumn{1}{|l}{518.7460} & 0.495 & $-2.300$ & \\  
Ce II & 527.4229 & 1.044 & $ 0.130$ & \\
Ce II & 533.0556 & 0.869 & $-0.400$ & \\ 
Ce II & 604.3373 & 1.206 & $-0.480$ & \\ 
Ce II & 605.1815 & 0.232 & $-1.530$ & \\ 
Ce II & 871.6659 & 0.122 & $-1.980$ & \\ 
Ce II & 877.2135 & 0.357 & $-1.260$ & \\
Ce II & 891.0948 & 0.435 & $-1.800$ & \\
\hline
\end{longtable}
\end{longtab}

\twocolumn
%\onecolumn

\section{Radial-velocity curves}
\label{Sect:RV-fig}

\begin{figure}
\includegraphics[width=9cm]{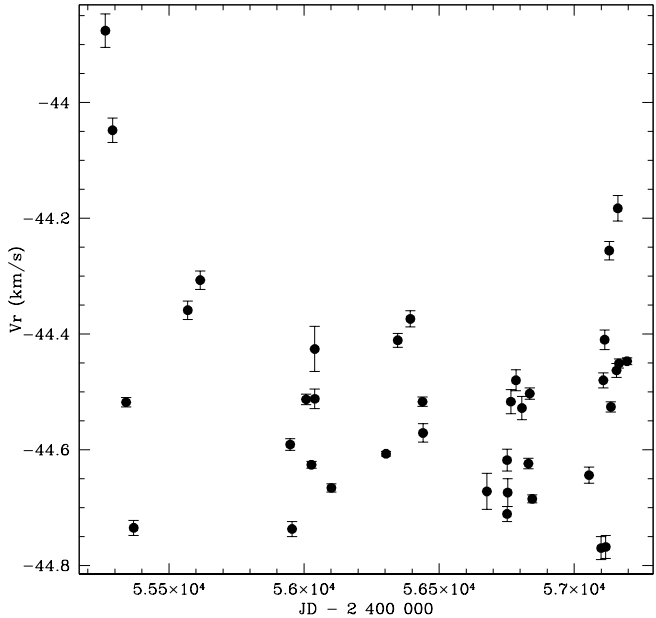}
\caption{\label{Fig:Vr1}
Radial-velocity data  for HE~1429-0551. 
}
\end{figure}

\begin{figure}
\includegraphics[width=9cm]{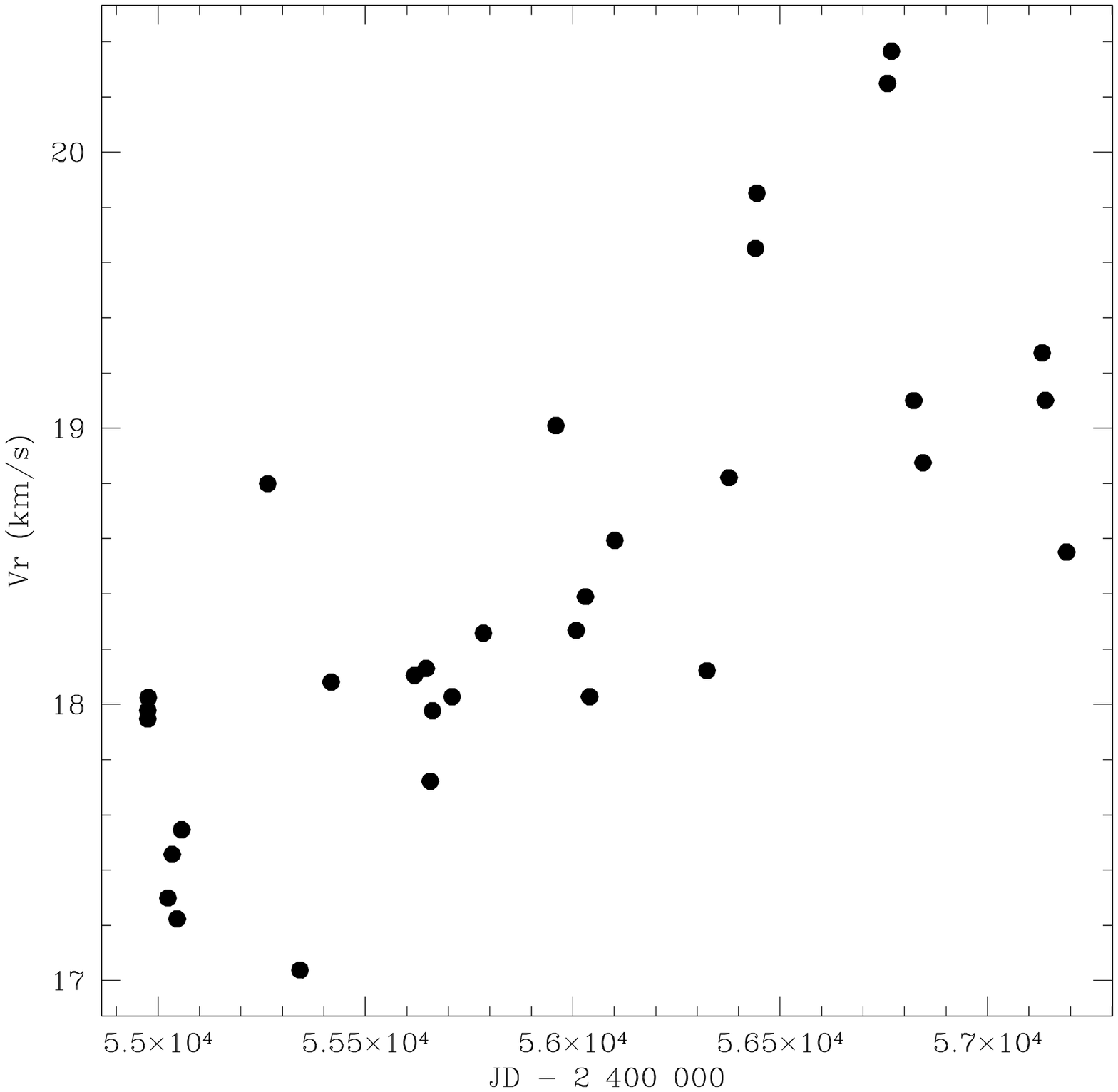}
\caption{\label{Fig:Vr2}
As Fig.~\ref{Fig:Vr1} for HD~145777. 
%The short-term variations are characterised by a (pseudo-?) Keplerian orbit 
%of period 326.7~d and eccentricity 0.33.
}
\end{figure}

\begin{figure}
\includegraphics[width=9cm]{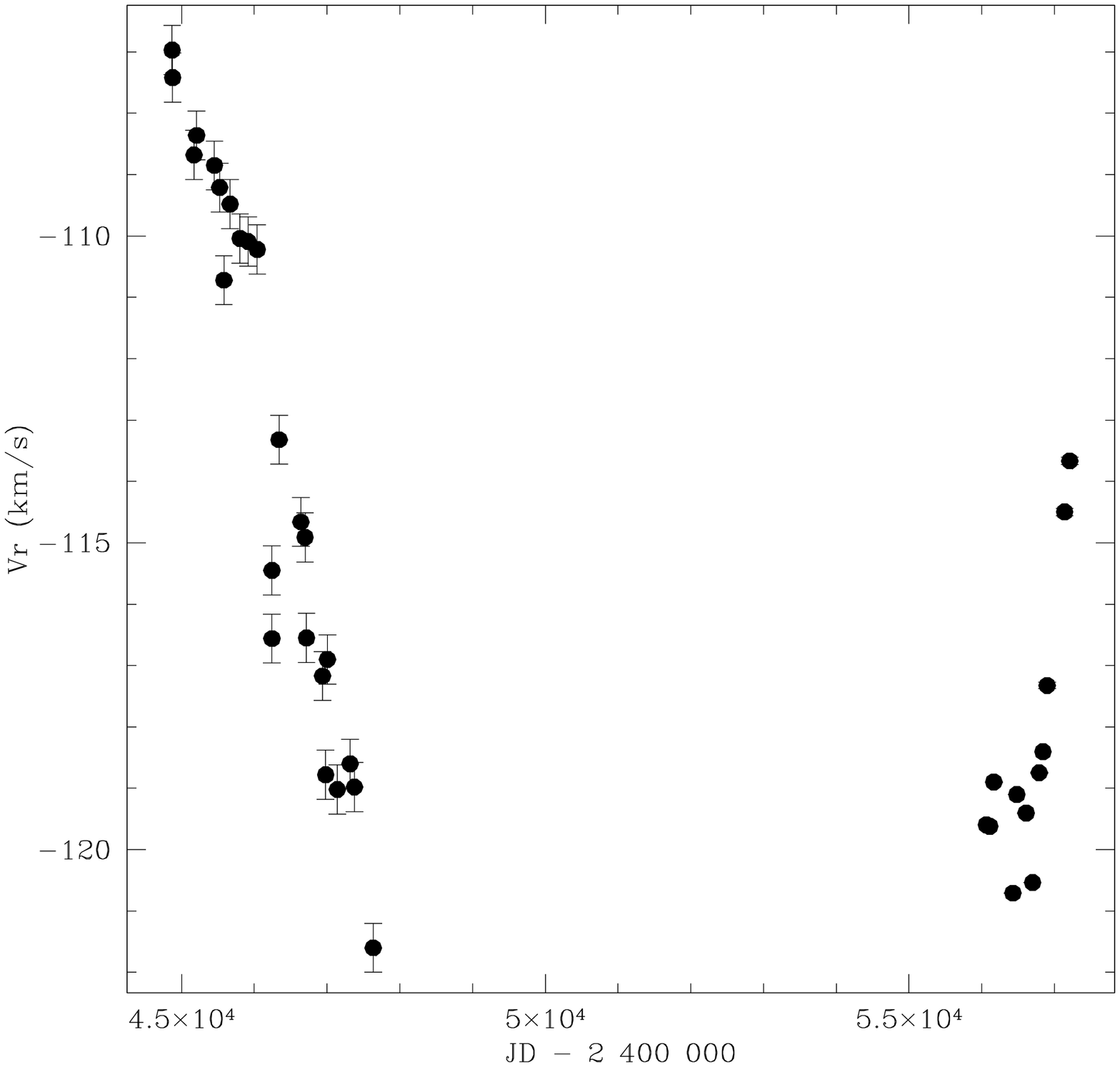}
\caption{\label{Fig:Vr3}
As Fig.~\ref{Fig:Vr1} for HD~187216, including data from  \citet{1990ApJ...352..709M}. 
}
\end{figure}

\begin{figure}
\includegraphics[width=9cm]{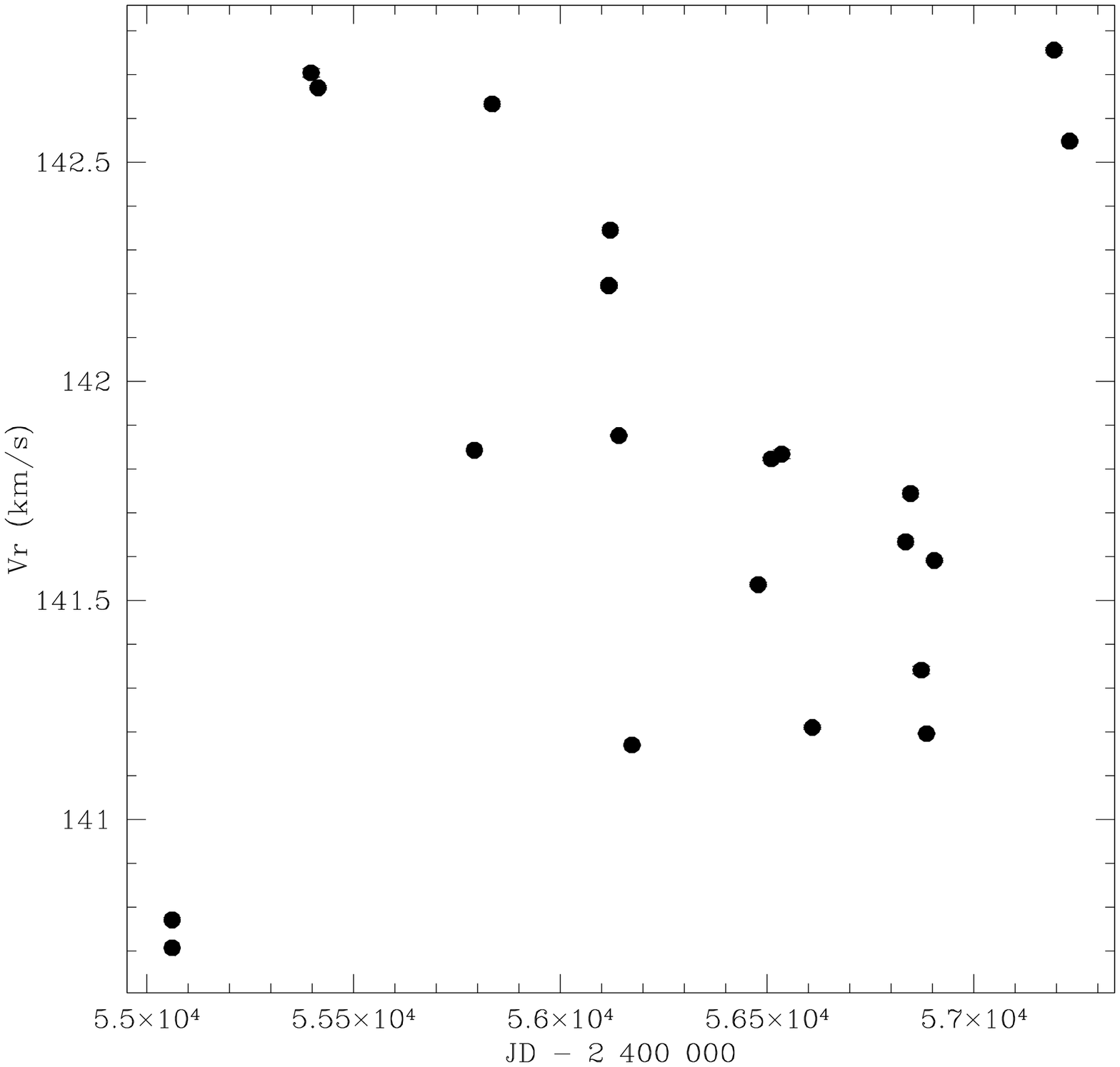}
\caption{\label{Fig:Vr4}
As Fig.~\ref{Fig:Vr1} for HE~2144-1832.
}
\end{figure}

\begin{figure}
\includegraphics[width=9cm]{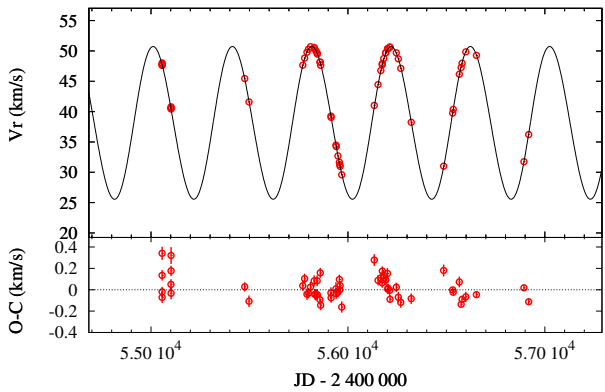}
\caption{\label{Fig:orbit1}
Orbital solution for HE~0111-1346. The lower panel shows the O-C deviations. 
}
\end{figure}

\begin{figure}
\includegraphics[width=9cm]{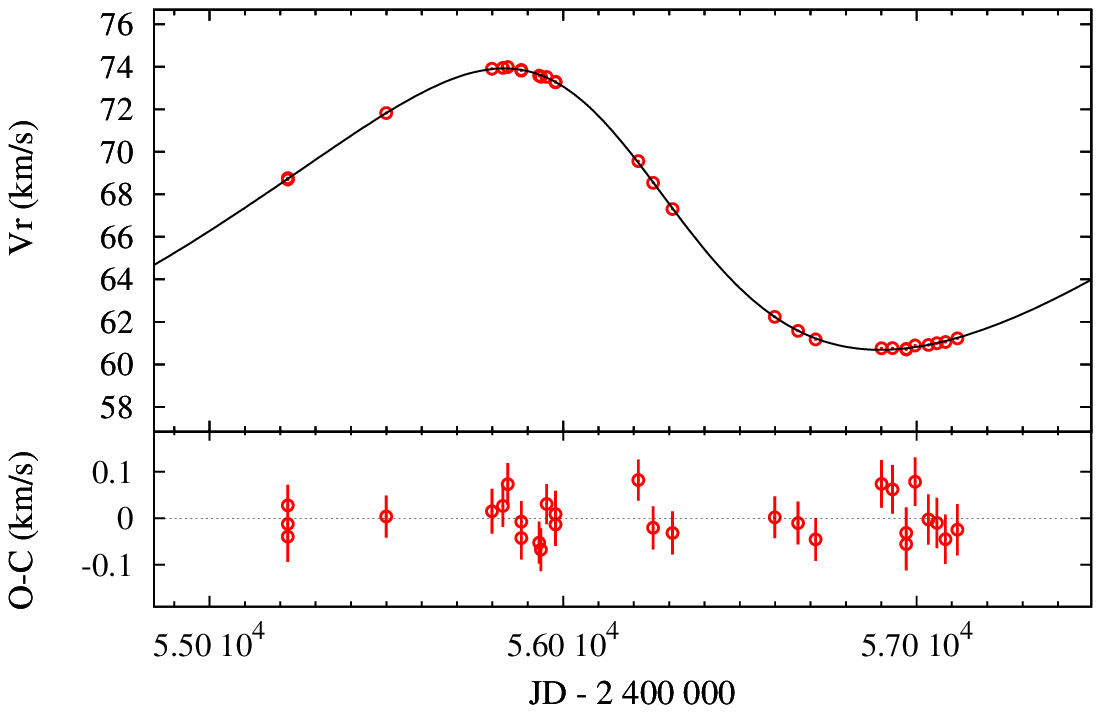}
\caption{\label{Fig:orbit2}
As Fig.~\ref{Fig:orbit1} for HE~0457-1805. 
}
\end{figure}

\begin{figure}
\includegraphics[width=9cm]{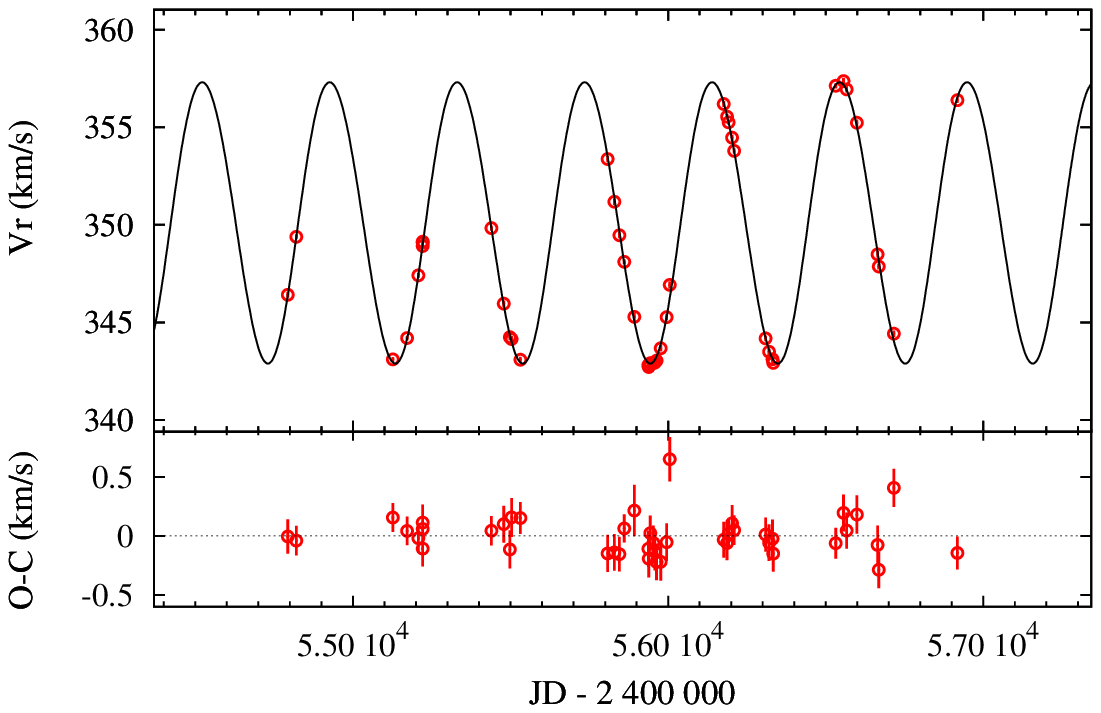}
\caption{\label{Fig:orbit3}
As Fig.~\ref{Fig:orbit1} for HE~0507-1653. 
}
\end{figure}

\begin{figure}
\includegraphics[width=9cm]{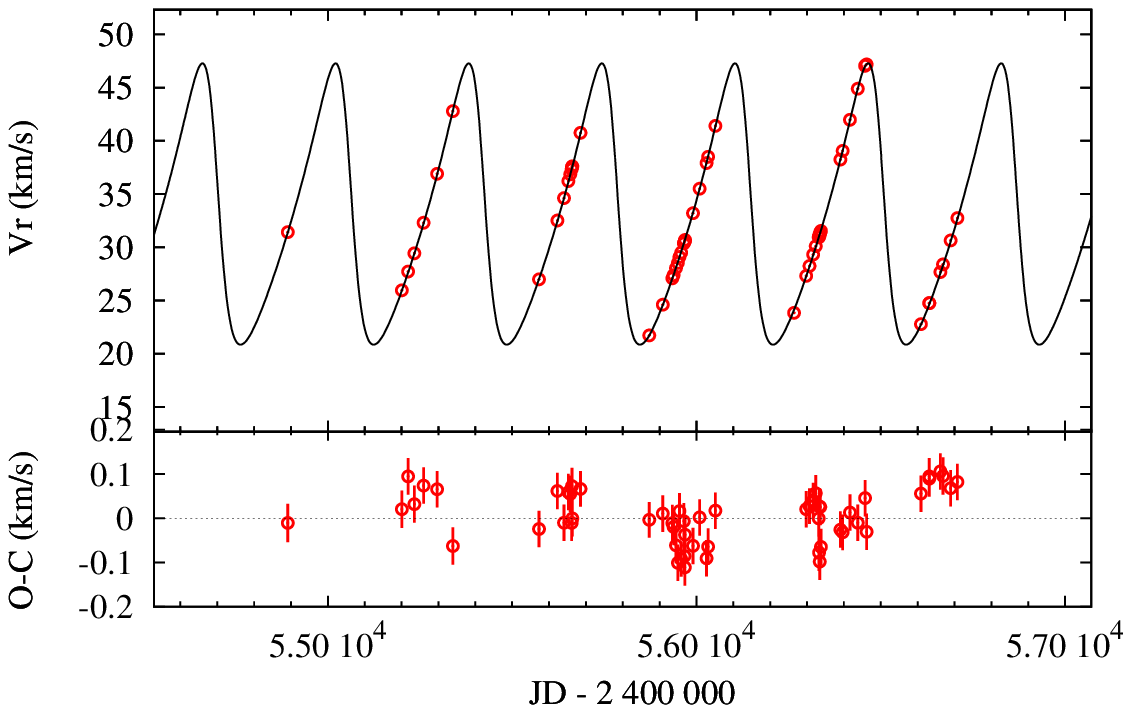}
\caption{\label{Fig:orbit4}
As Fig.~\ref{Fig:orbit1} for HIP~53522. 
}
\end{figure}

\begin{figure}
\includegraphics[width=9cm]{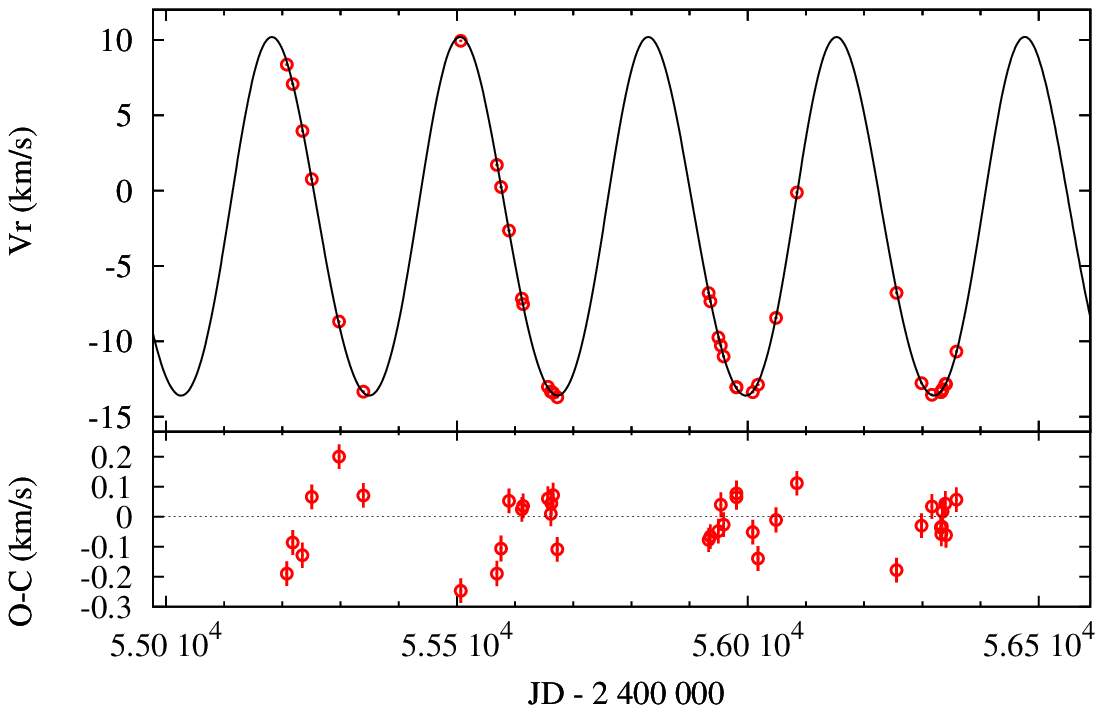}
\caption{\label{Fig:orbit5}
As Fig.~\ref{Fig:orbit1} for HIP~53832. 
}
\end{figure}

\begin{figure}
\includegraphics[width=9cm]{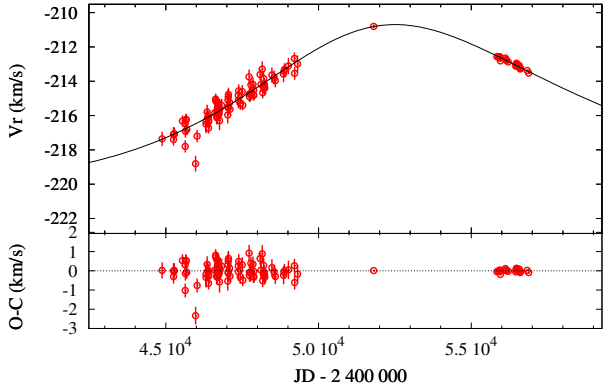}
\caption{\label{Fig:orbit6}
As Fig.~\ref{Fig:orbit1} for HD~26. The velocity point around JD~2\,452\,000 is from \citet{2003A&A...404..291V}, and the oldest measurements are from CORAVEL \citep{1979VA.....23..279B}. The HERMES and CORAVEL measurements are all on the IAU scale. The orbital elements are not yet fully constrained (the orbital period could be longer).
}
\end{figure}

\begin{figure}
\includegraphics[width=9cm]{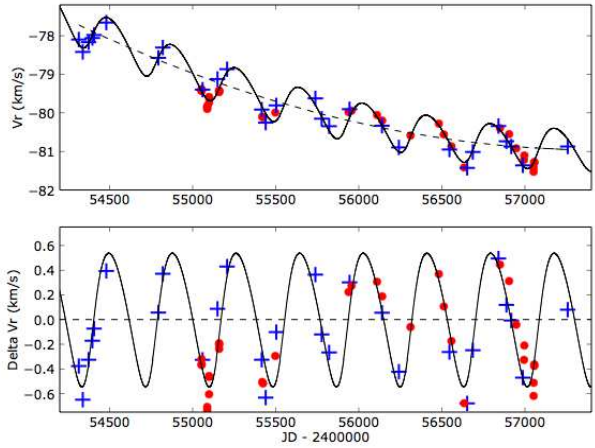}
\caption{\label{Fig:jitter1}
As Fig.~\ref{Fig:Vr1} for HE~0017+0055,  showing the short- and long-term orbital fits combined (top) as well as the short-term orbit alone \citep[red dots: HERMES velocities; blue plusses: Nordic Optical Telescope velocities; for details, see ][ and Table~\protect\ref{Tab:orbits_OC}]{2015A&A...XXX..NNNH}. 
}
\end{figure}

\begin{figure}
\includegraphics[width=9cm]{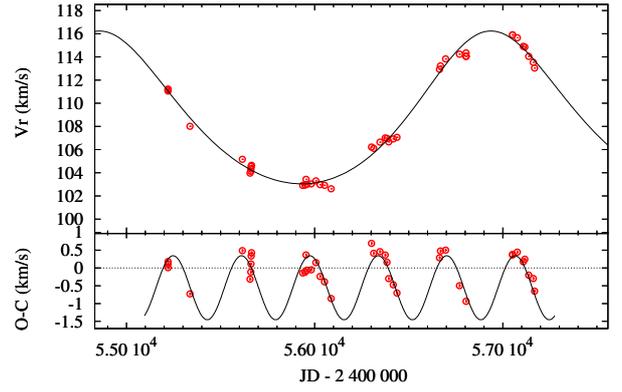}
\caption{\label{Fig:jitter2}
As Fig.~\ref{Fig:orbit1} for HE~1120-2122. The lower panel shows the low-amplitude, short-period variations through the $O-C$ residuals (see Table~\ref{Tab:orbits_OC}).
}
\end{figure}

\begin{figure}
\includegraphics[width=9cm]{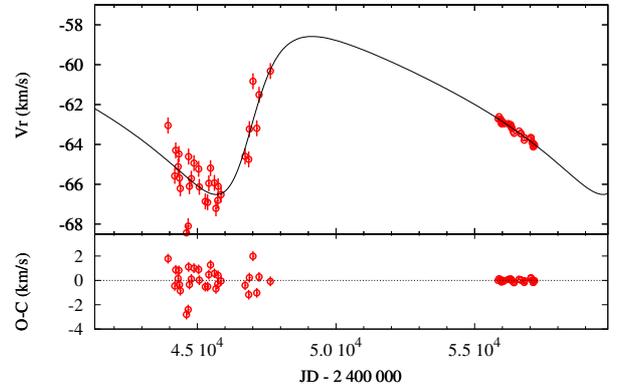}
\includegraphics[width=9cm]{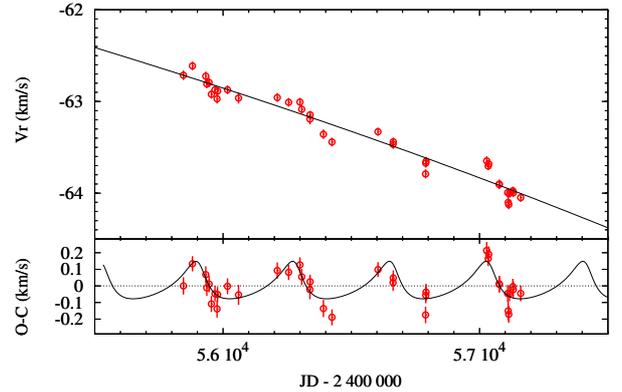}
\caption{\label{Fig:jitter3}
Top panel: As Fig.~\ref{Fig:orbit1} for HD~76396. Bottom panel: Zoom on the HERMES data, with the $O-C$ showing periodic variations (see Table~\ref{Tab:orbits_OC}).
}
\end{figure}

This appendix collects all figures presenting radial-velocity curves and orbits, listed in the same order as in Sect.~\ref{Sect:results}: SB? (Fig.~\ref{Fig:Vr1}), SB (Figs.~\ref{Fig:Vr2}--\ref{Fig:Vr4}), ORB (i.e., orbits with no significant residuals: Figs.~\ref{Fig:orbit1}--\ref{Fig:orbit6}), and finally SB+jitter or ORB+jitter (i.e., long-term trend or orbit with superimposed short-term variations:  Figs.~\ref{Fig:jitter1}--\ref{Fig:jitter3}). Individual radial velocities are available on-line from the {\it Centre de Donn\'ees Stellaires} (CDS).

HE~0507-1653 with its high radial velocity (350.2~\kms; Fig.~\ref{Fig:orbit3}) is particularly remarkable. Its proper motion has been estimated from the positions listed in the 2MASS 
catalogue (J05091656-1650046, epoch 1998.96, ICRS 2000: $\alpha = 77.319009, \delta = -16.834616$) and USNO-B catalogue (0731-0088818, epoch 1950, ICRS 2000: $\alpha = 77.318951, \delta = -16.834659$), namely $\mu_{\alpha} = 4.3$~mas~yr$^{-1}$, $\mu_{\delta} = 3.1$~mas~yr$^{-1}$. This proper motion
is not especially large, and should not add a substantial contribution to the large radial velocity. An accurate space velocity cannot be computed, though, in the absence of a distance estimate for that star.


%\begin{figure}
%\includegraphics[height=9cm,angle=270]{10636_OC.ps}
%\caption{\label{Fig:orbit8}
%As Fig.~\ref{Fig:orbit1} for HD~10636. 
%}
%\end{figure}

\end{document}